\documentclass[aps,prd,eqsecnum,showpacs,floatfix]{revtex4}
\usepackage{latexsym}
\usepackage{graphicx,subfigure}
\begin{document}

\thispagestyle{empty}

\title{
Super-horizon primordial black holes}
\author{$^{1,2}$Tomohiro Harada\footnote{Electronic
address:harada@scphys.kyoto-u.ac.jp}
and
$^{1}$B.~J.~Carr\footnote{Electronic
address:B.J.Carr@qmul.ac.uk}}
\affiliation{
$^{1}$Astronomy Unit, School of Mathematical Sciences, 
Queen Mary, University of London, 
Mile End Road, London E1 4NS, UK \\
$^{2}$Department of Physics, Kyoto University, Kyoto 606-8502, Japan}
\date{\today}

\begin{abstract}                
We discuss a new class of solutions to the Einstein equations which
 describe a primordial black hole (PBH) in a flat Friedmann background. 
Such solutions
 arise if a Schwarzschild black hole is patched onto a Friedmann
 background via a transition region. They are possible 
 providing the black hole event horizon
 is larger than the cosmological apparent horizon. 
Such solutions have a number of strange features.  In particular, one
 has to define the black hole and cosmological horizons carefully
and one then finds that the mass contained within the
black hole event horizon decreases 
when it is larger than the Friedmann 
cosmological apparent horizon, although 
its area always increases. 
These solutions involve two distinct future null 
infinities and are interpreted 
as the conversion of a white hole 
into a black hole.
Although such solutions may not form from gravitational collapse in the
 same way as standard PBHs,
there is nothing unphysical about them, since
all energy and causality conditions are satisfied.
Their conformal diagram is a natural amalgamation of the Kruskal diagram for the extended Schwarzschild solution and the conformal diagram for a black hole in a flat Friedmann background.
In this paper, such solutions are obtained numerically for a
spherically symmetric universe containing a massless scalar field, but 
it is likely that they exist for more general matter fields and 
less symmetric systems.
\end{abstract}
\maketitle

\section{Introduction}

In a recent paper~\cite{hc2004b} 
(henceforth Paper I) we studied numerically the
growth of primordial black holes (PBHs) in a universe containing
a massless scalar field but no other matter.
Following Hamad\'e and Stewart~\cite{hs1996},  
the double-null formulation of the Einstein equations was used.
This is a powerful tool for investigating 
regions outside the cosmological horizon and 
inside the black hole horizon 
simultaneously. On the assumption that the PBH is formed from a local initial density perturbation which
propagates causally, the black hole was modelled by matching a Schwarzschild solution to an exact flat Friedmann solution across a null surface. Initial data were specified on an
outgoing and ingoing null surface. They were assumed to be exactly Friedmann on the outgoing surface and outside the matching boundary on the ingoing surface
but some perturbation of Friedmann inside the matching boundary.

In all the solutions considered in Paper I, the black hole 
event horizon (BHEH) was assumed to be smaller than 
the cosmological apparent horizon.
However, in some circumstances (e.g. in the inflationary scenario), the
perturbation - and perhaps even the black hole itself - may extend well beyond it. In this
case, the only upper limit on its size comes from
requiring that the perturbed region be part of our universe rather than a
separate closed universe. The condition for this has been derived
precisely for the situation in which the collapsing region is
homogeneous and the equation of state is $p=k\rho$~\cite{hc2004a}. In
particular, this includes the massless scalar field case, since this is
equivalent to a stiff fluid with $k=1$ providing the gradient of the
field is everywhere timelike and vorticity free (as expected).
The peculiarity of the formation of 
PBHs in a universe with a stiff fluid 
was discussed in Ref.~\cite{zn1980}, and this suggests that the 
gradient of the scalar field could have an important hydrodynamical effect.

We study such ``super-horizon'' solutions in this paper, although our numerical results only cover the scalar field case. 
It turns out that these scalar field solutions have some rather strange properties. For example, the mass contained within the 
BHEH decreases when the PBH is larger than the 
cosmological apparent horizon, although 
the area always increases. 
Since our system satisfies all energy conditions,
the mass decrease of super-horizon PBHs we show here 
is completely different from the black hole mass 
decrease due to phantom energy accretion~\cite{bde2004}.
Also the conformal diagram for these solutions is interesting, being a
natural extension of the Kruskal diagram to the cosmological context.
The scalar field usually has a timelike gradient in the numerical
simulations shown here and this means that it is equivalent to 
a stiff fluid.
It is likely that there are analogues of these solutions for
more general fluids with $p=k\rho$ providing $k>1/3$. 
However, we do not study these solutions here.
In all these cases, we need to define cosmological and black hole
horizons very carefully when their sizes are all of the same order.

\section{Formulation}
\subsection{Double-Null Formulation of Einstein equations}
\label{sub:double-null}
As in paper I, we consider a massless scalar field $\Psi$ in 
general relativity, for which 
the stress-energy tensor is
\begin{equation}
T_{ab}=\Psi_{,a}\Psi_{,b}-\frac{1}{2}g_{ab}\Psi^{,c}\Psi_{,c}.
\end{equation}
The Einstein equations are
\begin{equation}
R_{ab}-\frac{1}{2}g_{ab}R=8\pi T_{ab},
\label{eq:einstein}
\end{equation}
and the equation of motion for the scalar field is
\begin{equation}
\Box\Psi=\Psi^{;a}_{~~;a}=0.
\label{eq:eom}
\end{equation}
We focus on a spherically symmetric system,
for which the line element can be written in the form
\begin{equation}
ds^{2}=-a^{2}(u,v)dudv+r^{2}(u,v)(d\theta^{2}+\sin^{2}\theta d\phi^{2}),
\end{equation}
where $u$ and $v$ are advanced and retarded time coordinates,
respectively, $a$ is the metric function
and $r$ is the ``area radius'' (the proper area of the sphere
of constant $r$ being $4\pi r^2$).
Equations~(\ref{eq:einstein}) and (\ref{eq:eom}) then imply that
we have 14 first-order partial differential equations
and two auxiliary equations. 
These equations are given explicitly in Section 2 of \cite{hs1996}.
We adopt units in which $G=c=1$.

\subsection{Misner-Sharp mass and trapping horizons}
\label{sec:mass_horizon}
The existence and position of marginal surfaces 
can be inferred from the form of the 
Misner-Sharp mass~\cite{ms1964}. This is a well-behaved
quasi-local mass in spherically symmetric 
spacetimes~\cite{hayward1996}, which can be written as
\begin{equation}
m=\frac{r}{2}\left(1+\frac{4r_{,u}r_{,v}}{a^{2}}\right).
\label{eq:misner_sharp}
\end{equation}
Combining this with the equations in Section 2 of \cite{hs1996},
we can derive the following useful relations:
\begin{eqnarray}
m_{,u}&=&-\frac{8\pi r^{2}r_{,v}(\Psi_{,u})^{2}}{a^{2}} ,
\label{eq:m_u}\\
m_{,v}&=&-\frac{8\pi r^{2} r_{,u}(\Psi_{,v})^{2}}{a^{2}}.
\label{eq:m_v}
\end{eqnarray}

In this paper, we adopt the ``trapping horizon''
framework introduced by Hayward~\cite{hayward1993,hayward1996}
because this provides a systematic and mathematically 
transparent view of the sort of
cosmological black holes treated here.
In the context of this paper,
trapping horizons and conventional apparent horizons are almost
equivalent but they need not be in more general situations.

Although the values for $r_{,v}$ and $r_{,u}$
are not geometrical invariants, their signs are, which leads to the following definitions.
(i) A metric sphere is said to be trapped if $r_{,v}r_{,u}>0$. 
A sphere with $r_{,u}<0$ and $r_{,v}<0$ is future trapped,
while one with $r_{,u}>0$ and $r_{,v}>0$ is past trapped.
(ii) A metric sphere is said to be untrapped if  $r_{,v}r_{,u}<0$.
On an untrapped surface, $\partial_{v}$ 
is outgoing if $r_{,v}>0$ and ingoing if $r_{,v}<0$.
More generally, a spacelike or null normal 
vector $z^{a}$ is outgoing if $z^{a}r_{,a}>0$
and ingoing if $z^{a}r_{,a}<0$.
(iii) A metric sphere is said to be marginal if  $r_{,v}r_{,u}=0$.
A sphere with $r_{,v}=0$ is future marginal
if $r_{,u}<0$, past marginal
if $r_{,u}>0$ and bifurcating marginal if $r_{,u}=0$.
It is described as outer marginal if $r_{,uv}<0$,
inner marginal if $r_{,uv}>0$ and degenerate marginal if $r_{,uv}=0$. 
It is easily seen that a sphere is marginal if and only 
if $r=2m$, trapped if and only if $r<2m$ and untrapped if and only if 
$r>2m$. 

The closure of a hypersurface foliated by a future or past, outer or 
inner marginal sphere is called a (nondegenerate) trapping horizon.
In Hayward's approach, the black hole apparent horizon
is replaced with a ``future outer trapping horizon'' (FOTH), while 
the cosmological (white hole) apparent horizon is
replaced with a ``past outer trapping horizon'' (POTH). 
There is a critical difference between an apparent horizon and 
a trapping horizon.
An apparent horizon is defined only on a prescribed spacelike
hypersurface, which is usually required to be a Cauchy surface, so it exists
only if the spacetime is strongly 
asymptotically predictable~\cite{wald1983}.
On the other hand, a trapping horizon is the trajectory 
of a marginal surface in the whole spacetime, and it is defined locally whenever a null foliation is possible. The spacetime does not 
need to be strongly asymptotically predictable.
It should be stressed that Hawking's 
``area theorem''~\cite{hawking1971b} only refers to the BHEH.
On the other hand, Hayward has shown that it also applies for a
FOTH providing the null energy condition holds. 

Providing the black hole horizon is within the cosmological horizon and $u$ and $v$
are the standard double-null coordinates 
in the asymptotic Friedmann region, 
the FOTH and the POTH correspond to the conditions 
$r_{,v}=0$ and $r_{,u}=0$, respectively.
However, the situation is more complicated if 
the black hole horizon is outside the cosmological horizon. In this
case, we can still define trapping horizons
by the conditions $r_{,v}=0$ and $r_{,u}=0$, but these are no longer everywhere 
identified with a FOTH
and a POTH. Using the above equations, we can show that along 
a trapping horizon, on which $r=2m$, we have
\begin{equation}
\left[a^{2}r_{,u}+16\pi r^{2}r_{,v}(\Psi_{,u})^{2}\right]du
+\left[a^{2}r_{,v}+16\pi r^{2}r_{,u}(\Psi_{,v})^{2}\right]dv
=0.
\end{equation}
On trapping horizons, which have $r_{,v}=0$ and $r_{,u}=0$, we therefore have 
\begin{eqnarray}
&&a^{2}du+16\pi r^{2}(\Psi_{,v})^{2}dv=0, \\
&&16\pi r^{2}(\Psi_{,u})^{2}du+a^{2}dv=0,
\end{eqnarray}
respectively.
We conclude that
trapping horizons are non-timelike in this system.
More precisely, a trapping horizon with $r_{,v}=0$ 
has $u=\mbox{const}$ if and only if $\Psi_{,v}=0$, while
a trapping horizon with $r_{,u}=0$
has $v=\mbox{const}$ if and only if $\Psi_{,u}=0$.
Except for these special cases, trapping horizons
are spacelike.

\begin{figure}[htbp]
\includegraphics[scale=1]{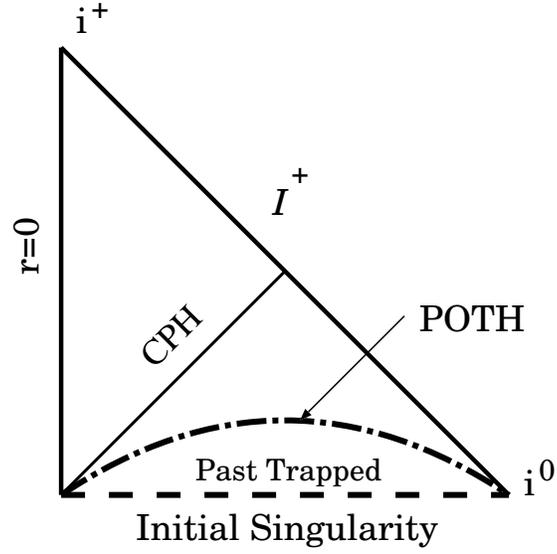}
\caption{\label{fig:flat_friedmann}
The conformal diagram for a flat Friedmann 
spacetime with a massless scalar field.
The cosmological apparent horizon, which is a 
past outer trapping horizon (POTH), 
is spacelike and outside 
the cosmological particle horizon (CPH).}
\end{figure}

The form of $a(u,v)$ and $r(u,v)$ in the exact flat Friedmann model 
is given in
Appendix B of Paper I. This implies that the cosmological particle
horizon has $u=0$, while the cosmological apparent horizon, which 
is a POTH, has $3u+v=0$.
This shows that the cosmological apparent horizon is spacelike and outside the particle
horizon. The conformal diagram of the spacetime is indicated in
Fig.~\ref{fig:flat_friedmann}, which shows the initial (big bang) spacelike
singularity. 
The cosmological apparent horizon also coincides with the Hubble horizon
in this case.
If one has a black hole embedded in an exact or asymptotically flat
Friedmann model and smaller than the cosmological particle horizon, 
the conformal diagram will change to the form indicated 
in Fig.~\ref{fig:pbh_clean}. This now contains a BHEH,
a FOTH and a final (black hole) spacelike singularity.

\section{Initial data for PBHs}
\subsection{Structure of initial data}
The initial data are prescribed on the two null surfaces $u=u_{0}$
and $v=v_{0}$, with 
the region of calculation being the diamond 
$[u_{0},u_{1}]\times [v_{0},v_{1}]$.
We have three independent functions
on the two null surfaces: $a$, $\Psi$ and $r$. Two of them can 
be chosen freely
and the third is determined by the constraint
equations on the null surfaces.
It is convenient to choose
\begin{equation}
a(u_{0},v),a(u,v_{0}),\Psi(u_{0},v),\Psi(u,v_{0})
\end{equation}
as the free initial data and to regard
\begin{equation}
r(u_{0},v),r(u,v_{0})
\end{equation}
as being determined
by the initial value equations.
We can regard $\Psi(u_{0},v)$ and $\Psi(u,v_{0})$ 
as the physical degrees of freedom in the initial data,
while the choice for $a(u_{0},v)$ and $a(u,v_{0})$ fixes the gauge.

In the flat Friedmann region, 
we adopt the coordinate system 
given in Appendix B1 of Paper I and
impose flat Friedmann initial data for $a$ and $\Psi$
on $u=u_{0}$ for $v_{0}\le v\le v_{1}$.
On $v=v_{0}$, we also choose flat Friedmann data 
for $a$ for $u_{0}\le u \le u_{1}$.
For $\Psi$, we use the same data 
on the initial null surface $v=v_{0}$
for $u_{0}\le u\le u_{\rm m}$, 
but $\Psi=\mbox{const}$ for $u_{\rm m}+\Delta u<u\le u_{1}$.
This is equivalent to Schwarzschild data
in coordinates penetrating the black hole.
The sudden transition from flat Friedmann data to 
Schwarzschild data results in a discontinuity 
at $u=u_{\rm m}$, which reduces the numerical 
accuracy. Hence we smooth the transition with some
smoothing length $\Delta u$; we use a quadratic function 
between $u_{\rm m}$ and $u_{\rm m}+\Delta u$,
so that $\Psi$ and $\Psi_{,u}$ are continuous.
More precisely, we impose the following initial data for $a$ and 
$s\equiv \sqrt{4\pi}\Psi$:
\begin{eqnarray}
a^{2}(u,v_{0})&=&C^{2}\left(\frac{u+v_{0}}{2}\right) 
\label{eq:A_on_v0},\\
s(u,v_{0})&=&
\left\{
\begin{array}{ll}
\displaystyle\frac{\sqrt{3}}{2}\ln\left(\displaystyle\frac{u+v_{0}}{2}\right)+s_{0} & \quad (u<u_{\rm m})\\
\displaystyle\frac{\sqrt{3}}{2}\left[\displaystyle\frac{(\Delta u)^{2}-(u_{\rm m}+\Delta u-u)^{2}}
{2\Delta u (u_{\rm m}+\Delta u)}+\ln\left(
\displaystyle\frac{u_{\rm m}+v_{0}}{2}\right)\right]+s_{0} &
\quad (u_{\rm m}\le u<u_{\rm m}+\Delta u) \\
\displaystyle\frac{\sqrt{3}}{2}\left[\displaystyle\frac{\Delta u}
{2(u_{\rm m}+v_{0})}+
\ln\left(\displaystyle\frac{u_{\rm m}+v_{0}}{2}\right)\right]+s_{0} &
\quad (u\ge u_{\rm m}+\Delta u)
\end{array}
\right. ,
\label{eq:s_on_v0}
\end{eqnarray}
on the initial null surface $v=v_{0}$,
and 
\begin{eqnarray}
a^{2}(u_{0},v)&=&C^{2}\left(\frac{u_{0}+v}{2}\right), \\
s(u_{0},v)&=&\frac{\sqrt{3}}{2}\ln\left(\frac{u_{0}+v}{2}\right)+s_{0},
\end{eqnarray}
on the initial null surface $u=u_{0}$.
Here $C$ and $s_0$ are constants and, without loss of generality, we can choose $C=1$ and $s_{0}=0$.
Note that although one has a Schwarzschild vacuum for $u\ge u_{\rm m}+\Delta u$, this situation only applies instantaneously at $v=v_0$, since the inflowing matter will fill this up immediately.

\begin{figure*}[htbp]
\subfigure[]{\includegraphics[scale=0.9]{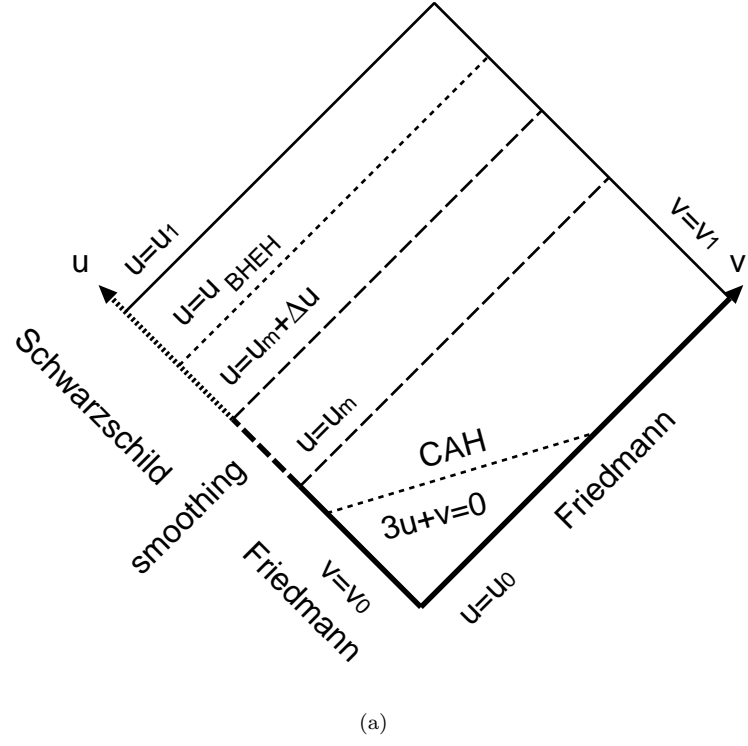}}
\subfigure[]{\includegraphics[scale=0.9]{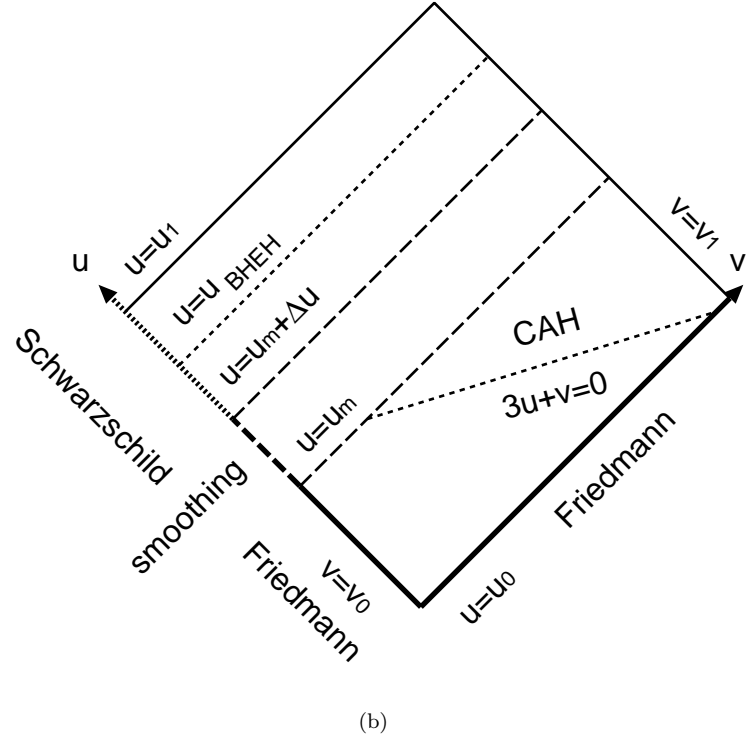}}
\caption{\label{fig:local_perturbation}
Schematic figures for the initial data
corresponding to PBHs (a) smaller and (b) larger than the cosmological
apparent horizon (CAH).}
\end{figure*}

Figure~\ref{fig:local_perturbation} depicts the initial data
for our numerical simulations, these being completely determined by the two parameters $u_{\rm m}$ and $\Delta u$. 
When the matching region is very narrow, the BHEH
is smaller than the Friedmann cosmological apparent horizon if $u_{\rm m}>u_{\rm CAH}(v_{0})$. This
corresponds to Fig.~\ref{fig:local_perturbation}(a), 
where $u=u_{\rm CAH}(v)$ denotes the trajectory of the 
cosmological apparent horizon. However,
if $u_{\rm m}<u_{\rm CAH}(v_{0})$, the BHEH is larger than the Friedmann cosmological horizon, which 
corresponds to Fig.~\ref{fig:local_perturbation}(b). 
Note that the cosmological apparent horizon would always go within the
matching radius if one took $v_0$ small enough, so when the black hole
forms is crucial. 
Note also that when we refer to a POTH in the background 
Friedmann model, we refer to it as the (Friedmann) cosmological
apparent horizon.
As discussed later, this relates to the distinction
between a black hole that forms from collapse and an eternal black hole
that exists ``ab initio''.

\section{Results}

\subsection{Black hole event horizon}
Our numerical code is based on that of Hamad\'e and Stewart~\cite{hs1996}
with a modification, described in Appendix A of Paper I, to ensure greater accuracy.
The initial data are prescribed on the two null surfaces $v=v_{0}=1$
and $u=u_{0}=-2/3$, this also fixing the units.
The calculated region is the diamond contained 
by $u=u_{0}$, $v=v_{0}$, $u=u_{1}$ and $v=v_{1}$. 
On the initial null surface $v=v_{0}$,
we make the matching at $u=u_{\rm m}$ and use the 
smoothing length $\Delta u$.
As time proceeds, the $u=\mbox{const}$ null rays will become ever
more sensitive to $r$ near the BHEH,
so the calculation 
is stopped at $v=v_{1}$, when the null rays become too coarse to be resolved.
The parameters used
are summarised in Table~\ref{tb:models} for six models. 
The BHEH is identified as the critical 
null ray $u=u_{\rm BHEH}$, the null rays with 
$u<u_{\rm BHEH}$
going to $r=\infty $ and those with $u>u_{\rm BHEH}$ 
returning to $r=0$. Therefore, the BHEH is only identified at the end of the calculation. 
Although the identification of the BHEH
is rather imprecise, because of numerical errors and the finiteness of the 
calculated regions,
it suffices for the physical interpretation of the 
results given below.
It is found that all six models
have a BHEH and Table~\ref{tb:models} indicates
the initial ratio of the sizes of the BHEH
and the Friedmann cosmological apparent horizon at $v=v_{0}$.
This complements Table I of Paper I, which only shows cases (A to D) in which the ratio is less than $1$.
All models except E have an initial BHEH
larger than the Friedmann cosmological apparent horizon.
In Model E the initial BHEH is
slightly smaller than the Friedmann cosmological apparent horizon. 

\begin{table}[htbp]
\begin{center}
\caption{\label{tb:models} Model parameters and
the initial mass ratios of BHEH to
the Friedmann cosmological apparent horizon.}

  \begin{tabular}{|c||c|c|c|c|c|c|c|}
   \hline
   Models & $u_{\rm m}$ & $\Delta u$ & $u_{0}$ & $u_{1}$ & $v_{0}$ & $v_{1}$
   & $m_{\rm BHEH}/m_{\rm CAH}$ \\
   \hline
   E & $-1/2$ & 0.4 & $-2/3$ & 2 & 1 & 4.5& 0.969\\
   F& $-1/2$ & 0.02 & $-2/3$ & 2 & 1 & 4.5 & 2.12\\
   G & $-1/2$ & 0.05 & $-2/3$ & 2 & 1 & 4.5 & 2.01\\
   H & $-1/2$ & 0.2 & $-2/3$ & 2 & 1 & 4.5& 1.52\\
   I & $-0.6$ & 0.02 & $-2/3$ & 2 & 1 & 4.5& 3.58\\
   J & $-0.4$ & 0.02 & $-2/3$ & 2 & 1 & 4.5& 1.31\\
\hline
  \end{tabular}
 \end{center}
\end{table}
\subsection{Location of horizons}
Figure~\ref{fig:horizons_uv} shows the locations 
of the BHEH and trapping horizons in the $(u,v)$ plane. 
It also indicates the signs of $r_{,v}$ and $r_{,u}$.
The region is future trapped if the signs are $(-,-)$, 
past trapped if they are $(+,+)$
and untrapped if they are $(+,-)$ or $(-,+)$.
Trapping horizons may have either
$r_{,v}=0$ or $r_{,u}=0$.

There are two qualitatively different 
classes of models.
In the first class, which includes all models except E, the initial BHEH
is larger than the Friedmann cosmological apparent horizon.
As seen from the figure, as $v$ increases from $v_{0}=1$,
two trapping horizons, one with $r_{,v}=0$
and the other with $r_{,u}=0$, appear and cross each other 
in the future of the BHEH.
In terms of the evolution with respect to $v$,
before the crossing the trapping horizons with $r_{,v}=0$ and $r_{,u}=0$
correspond to a POTH and a FOTH, respectively.
After the crossing, the situation is reversed, so
the trapping horizons with $r_{,v}=0$ and $r_{,u}=0$
correspond to a FOTH and a POTH, respectively. 
After further evolution, the POTH
crosses the BHEH. Thereafter it will coincide
with the Friedmann cosmological apparent horizon, 
while the FOTH corresponds to the black hole apparent horizon.

In the second class of models, which includes E, 
the initial BHEH is smaller than the 
Friedmann cosmological apparent horizon.
Trapping horizons with $r_{,v}=0$ and $r_{,u}=0$ appear and remain
in the future and past of the BHEH, respectively. These two trapping horizons do not cross each other
and always correspond to the FOTH and POTH, respectively.

\subsection{PBH mass change}
\begin{center}
\begin{figure*}[htbp]
\begin{tabular}{cc}
\subfigure[E]{\includegraphics[scale=0.5]{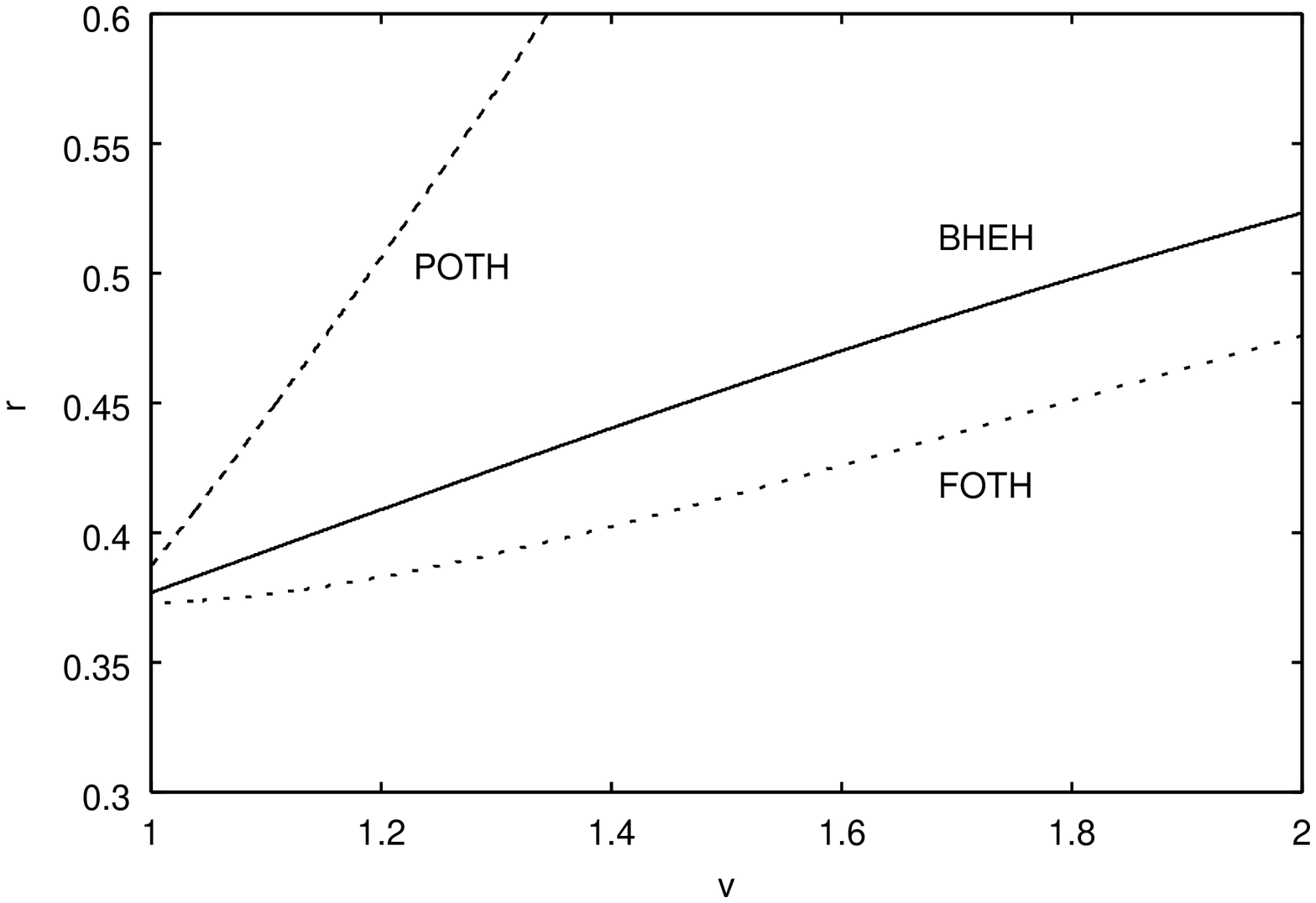}}&
\subfigure[F]{\includegraphics[scale=0.5]{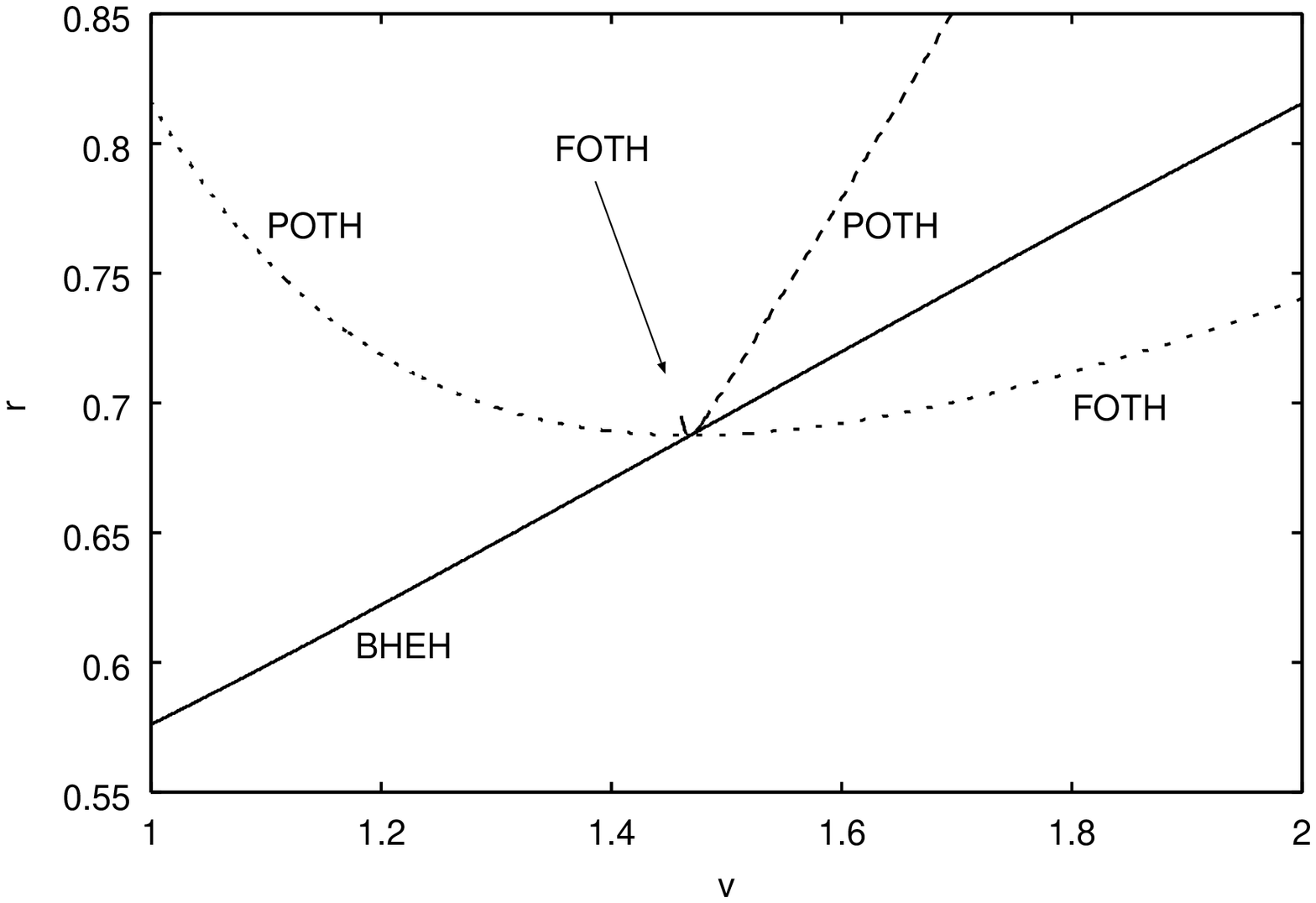}}\\
\subfigure[G]{\includegraphics[scale=0.5]{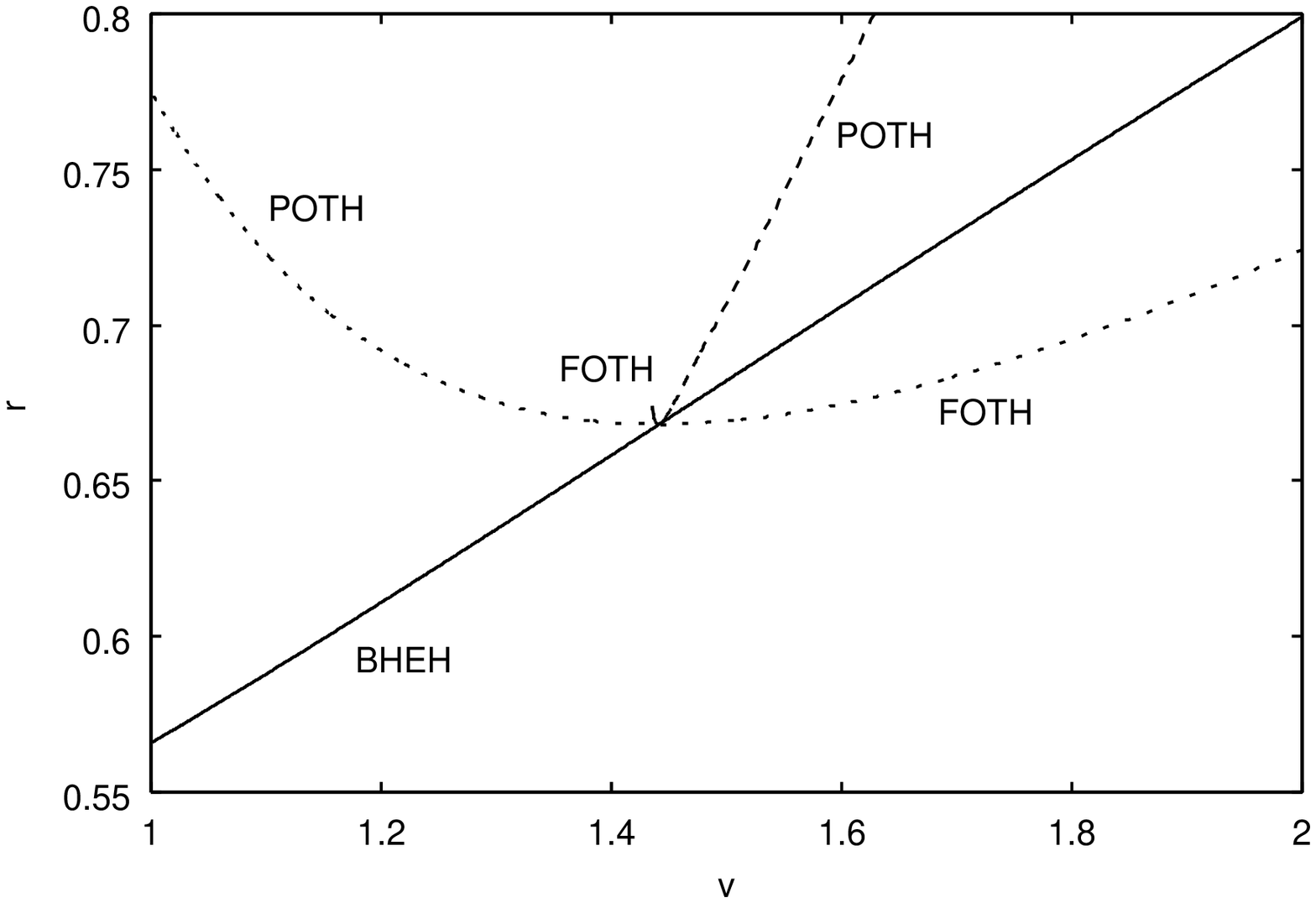}}&
\subfigure[H]{\includegraphics[scale=0.5]{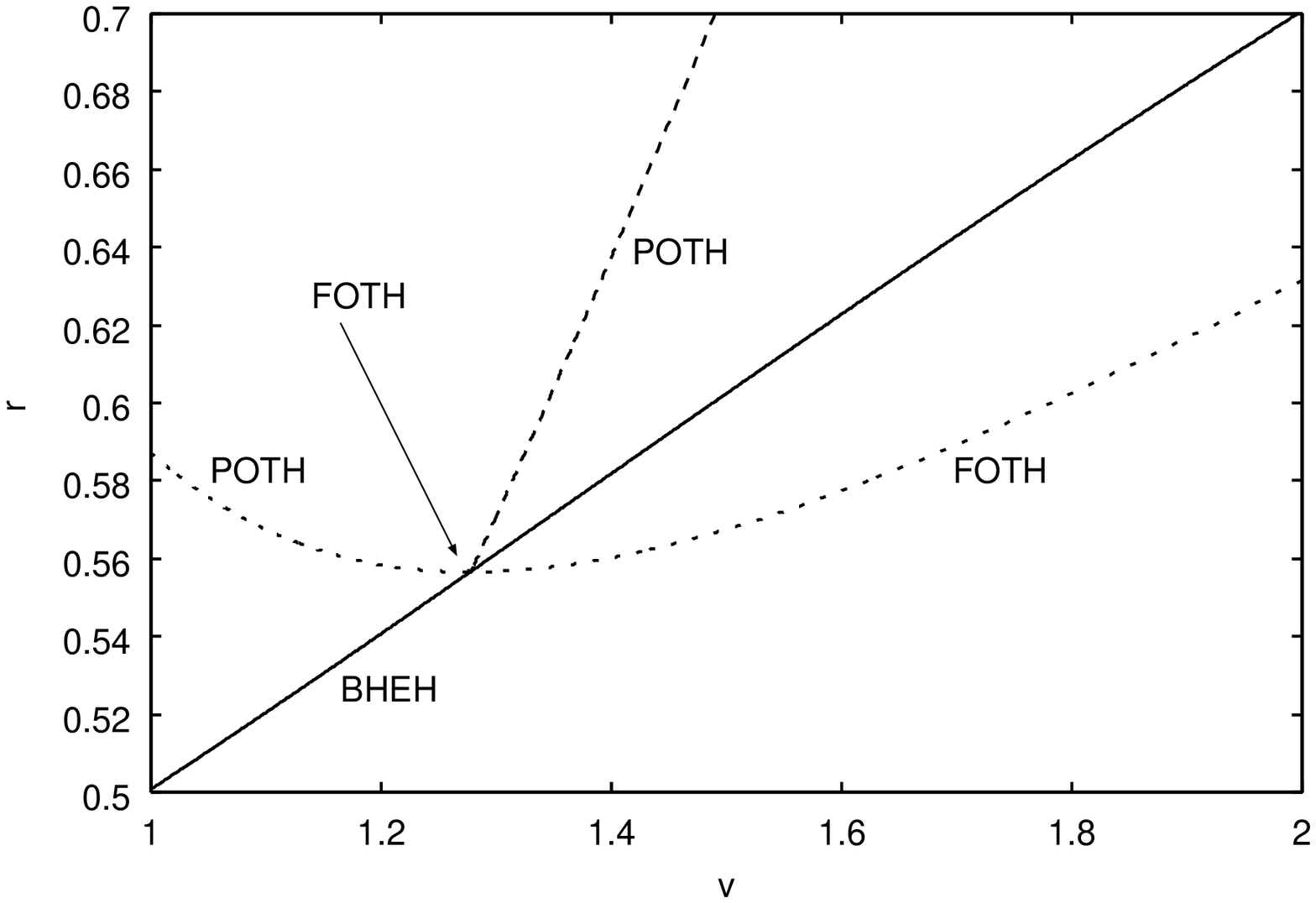}}\\
\subfigure[I]{\includegraphics[scale=0.5]{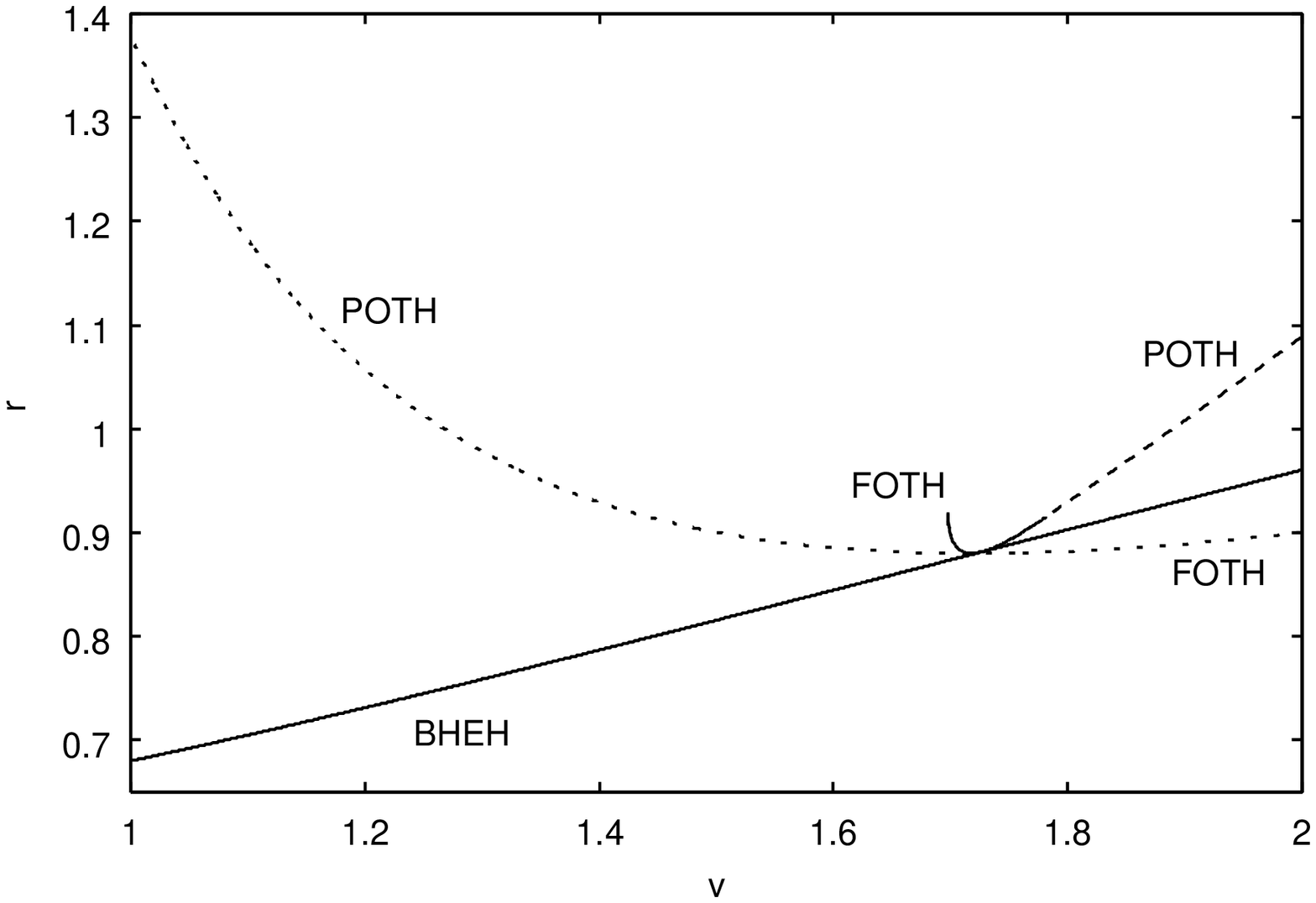}}&
\subfigure[J]{\includegraphics[scale=0.5]{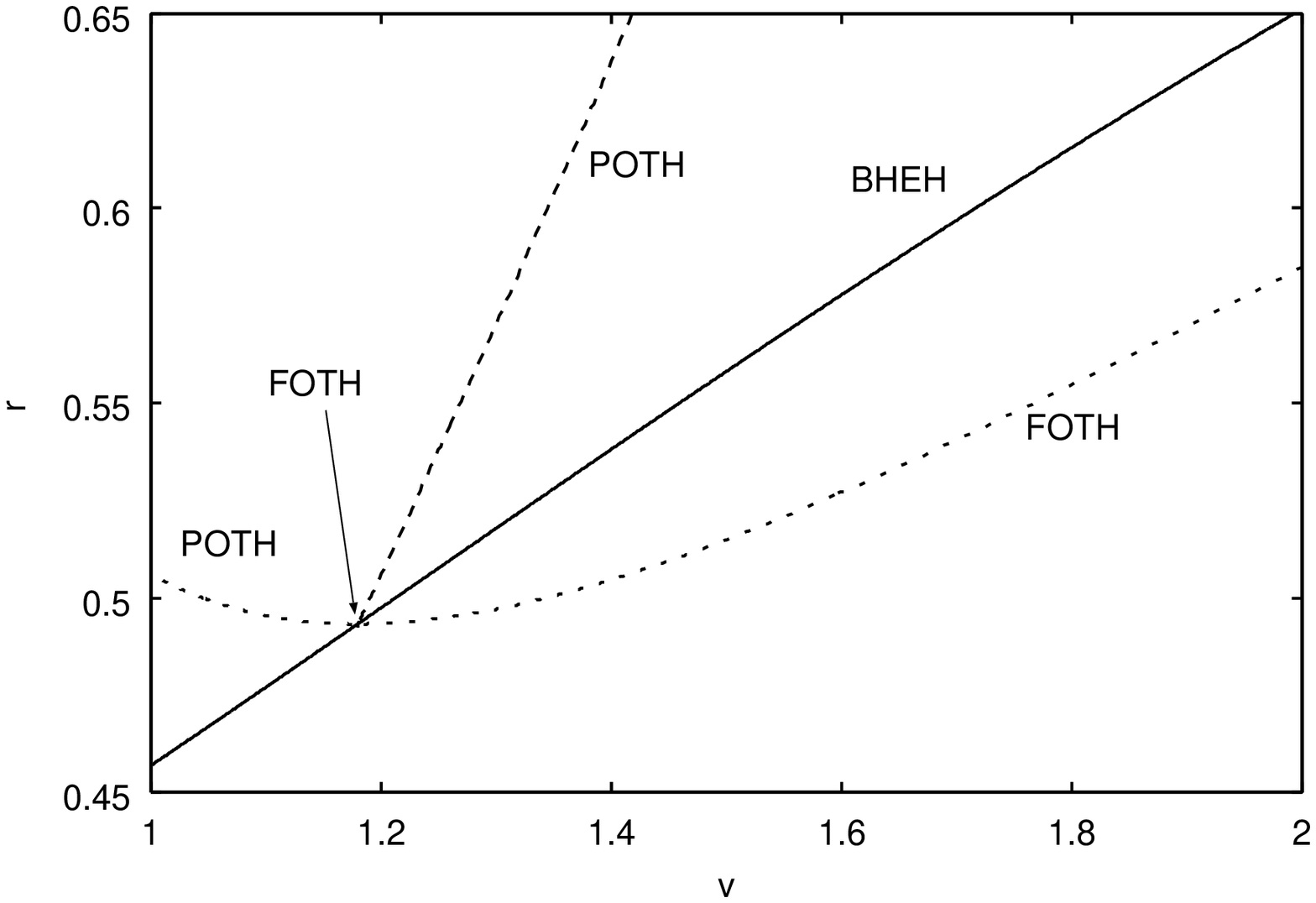}}
\end{tabular}
\caption{\label{fig:eh_rad}
The area radii of the black hole event horizon (BHEH) and
trapping horizons with $r_{,v}=0$
and $r_{,u}=0$, plotted with
solid, dotted and dashed lines, respectively. 
We can see the area radius of the black hole event horizon always
 increases.}
\end{figure*}
\end{center}

The area radii of the BHEH
and trapping horizons for these models are shown
in Fig.~\ref{fig:eh_rad}.
It is seen that the area of the BHEH
always increases for these models, which is 
consistent with the black hole area theorem~\cite{hawking1971b}.
If the initial BHEH is larger than 
the Friedmann cosmological apparent horizon,
i.e., for all models other than E, 
the areas of the trapping horizons with both $r_{,v}=0$ and $r_{,u}=0$ 
first decrease as $v$ increases and then increase after crossing each other.
It is interesting that, in terms of $v$,
the two trapping horizons cross
at a radius slightly larger than the BHEH at the crossing time.
After the BHEH enters the future of the POTH,
it soon gets much smaller
than it.
For these models, the FOTH appears just before the 
trapping horizons cross, although this is not so clear in the figure. 
For models where the initial BHEH is smaller than 
the Friedmann cosmological apparent horizon,
i.e., Model E, the situation is standard.
In terms of $r$, the FOTH is inside the BHEH,
while the POTH is outside it.
The BHEH soon gets much smaller than the POTH. 
 
\begin{center}
\begin{figure*}[htbp]
\begin{tabular}{cc}
\subfigure[E]{\includegraphics[scale=0.5]{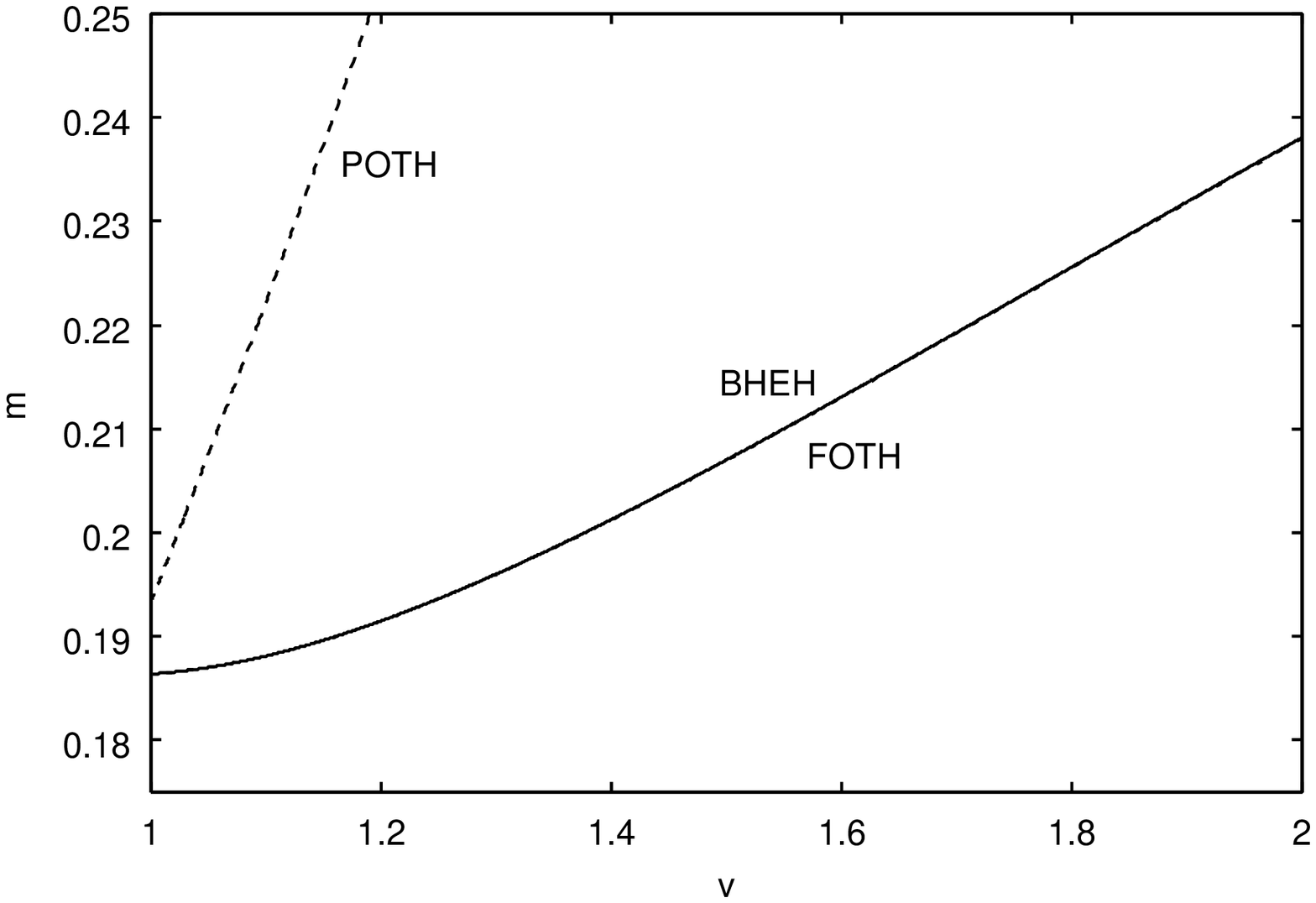}}&
\subfigure[F]{\includegraphics[scale=0.5]{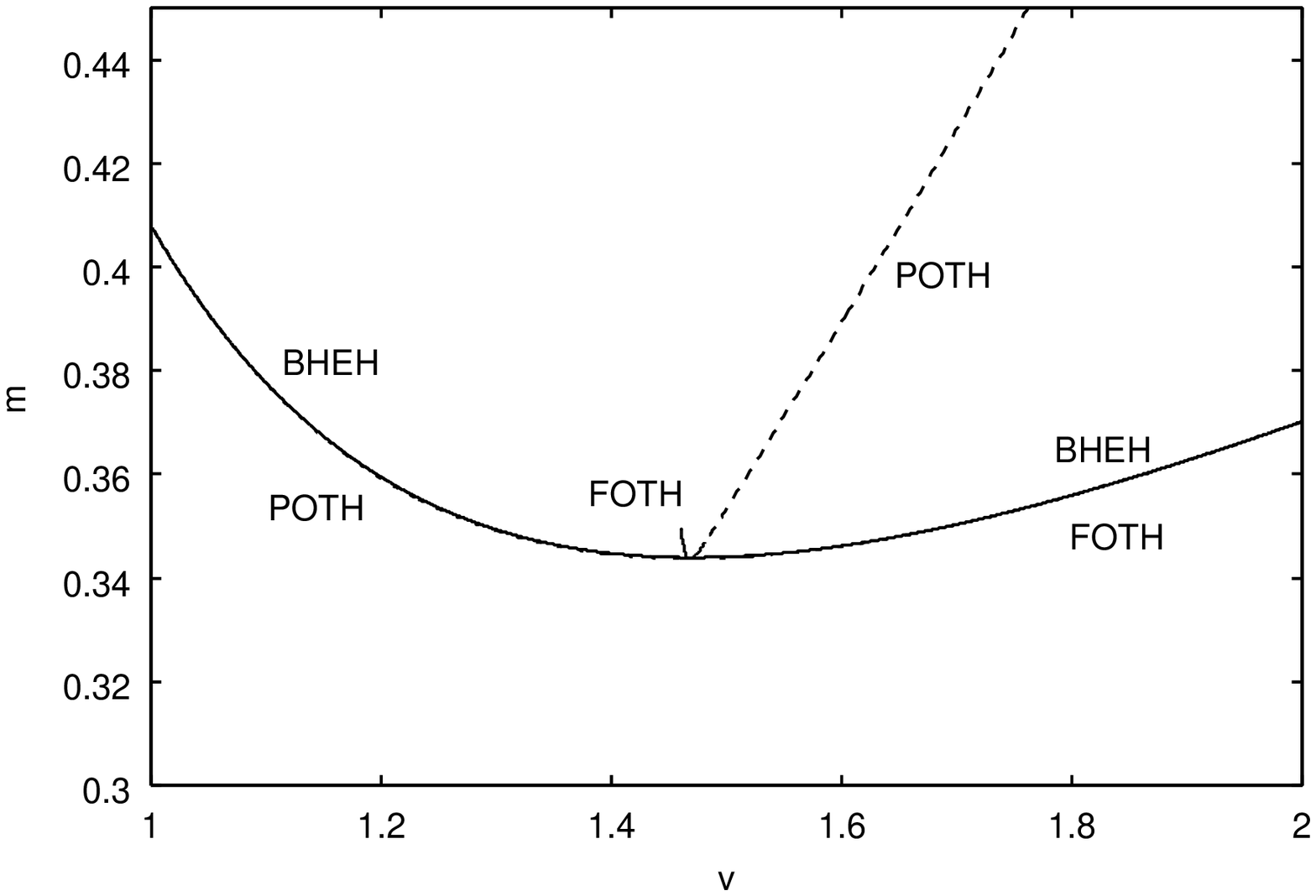}}\\
\subfigure[G]{\includegraphics[scale=0.5]{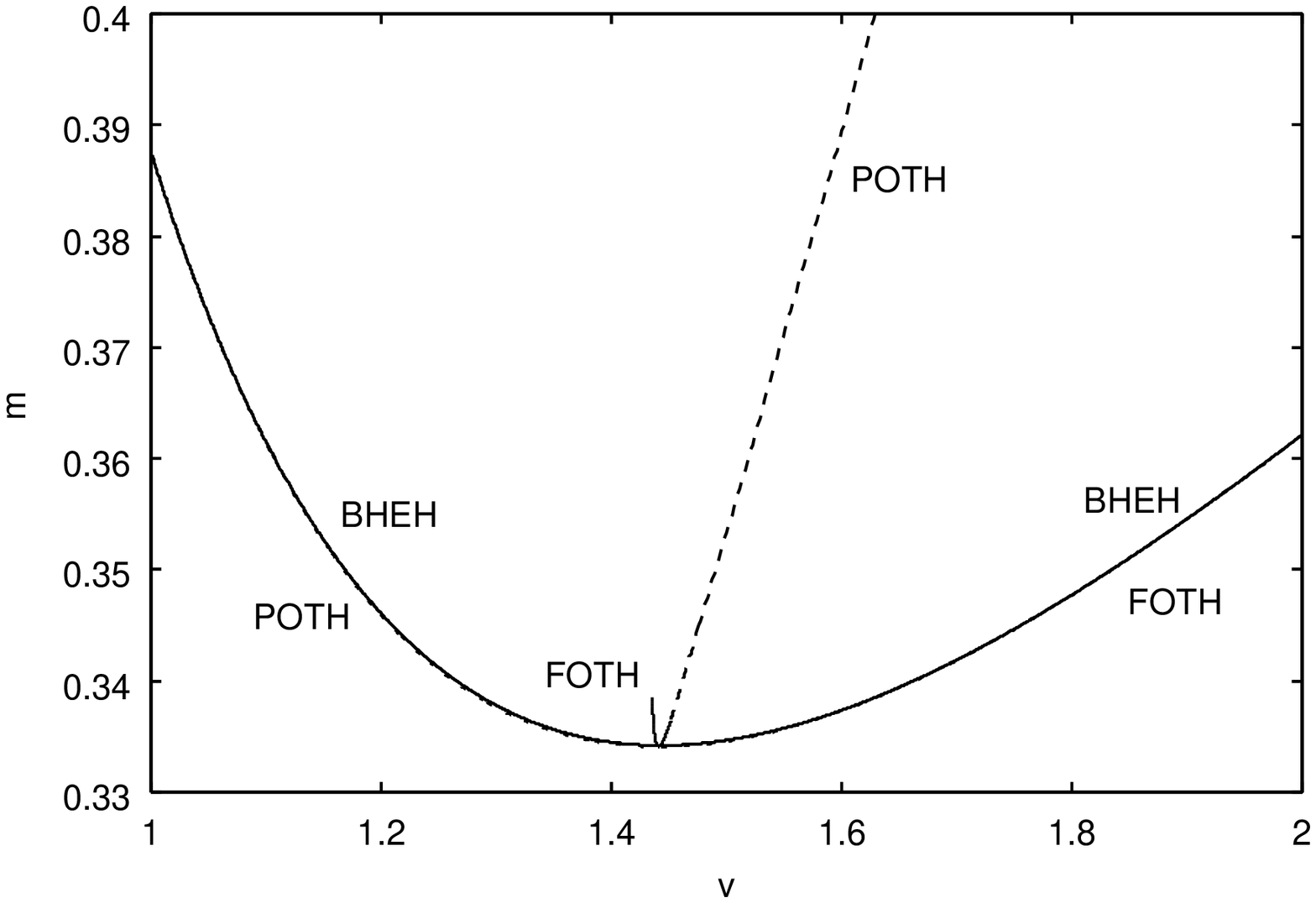}}&
\subfigure[H]{\includegraphics[scale=0.5]{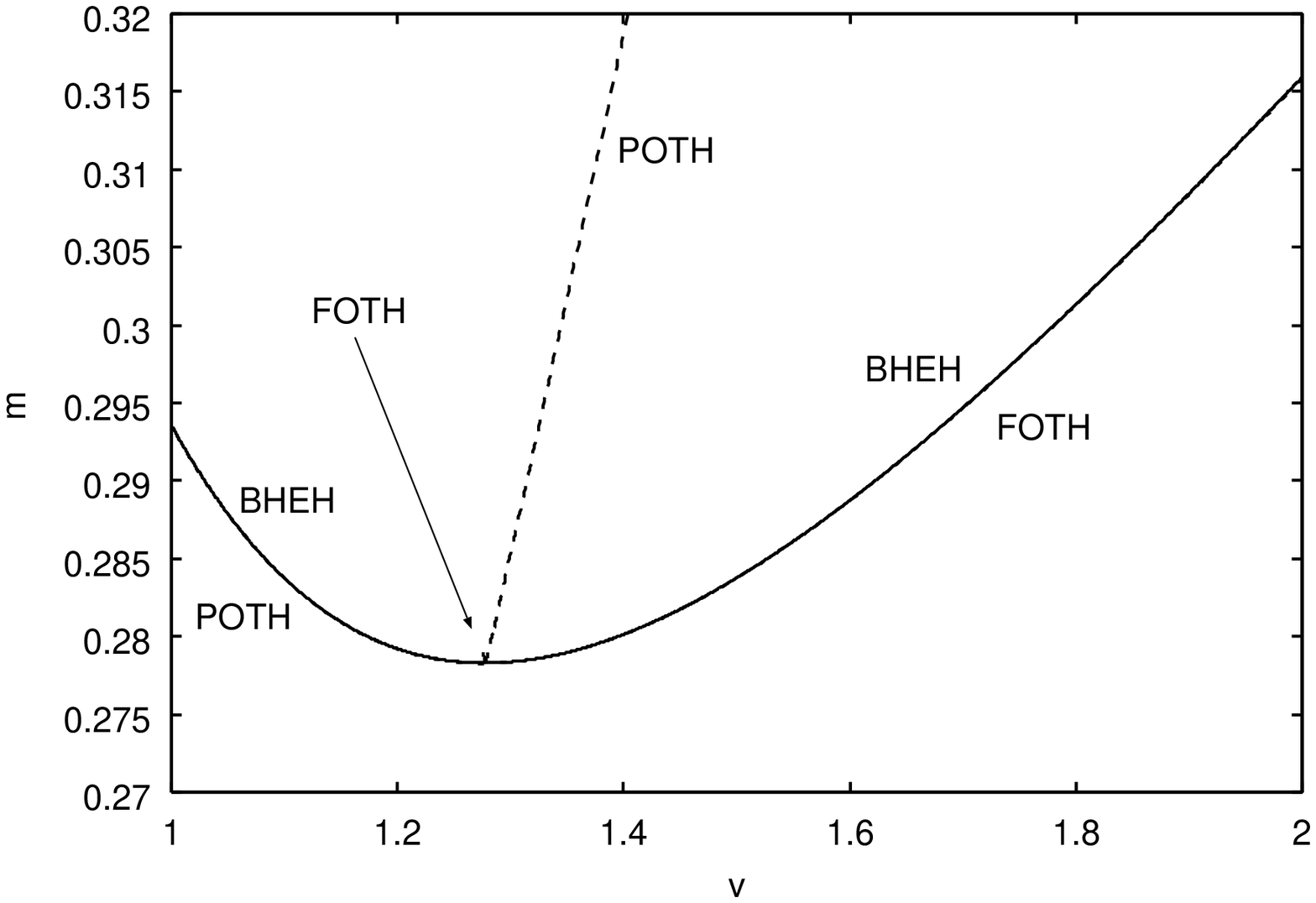}}\\
\subfigure[I]{\includegraphics[scale=0.5]{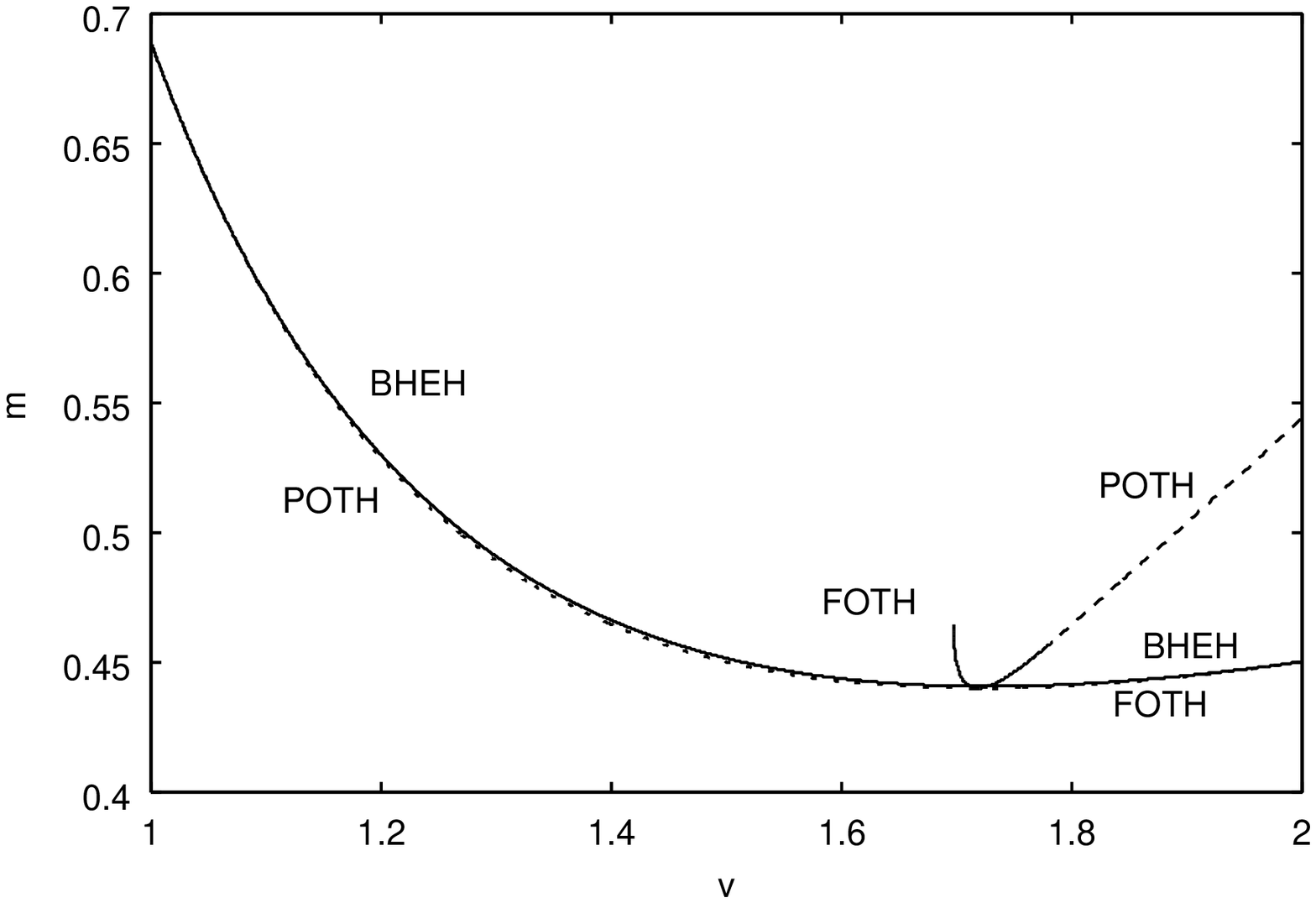}}&
\subfigure[J]{\includegraphics[scale=0.5]{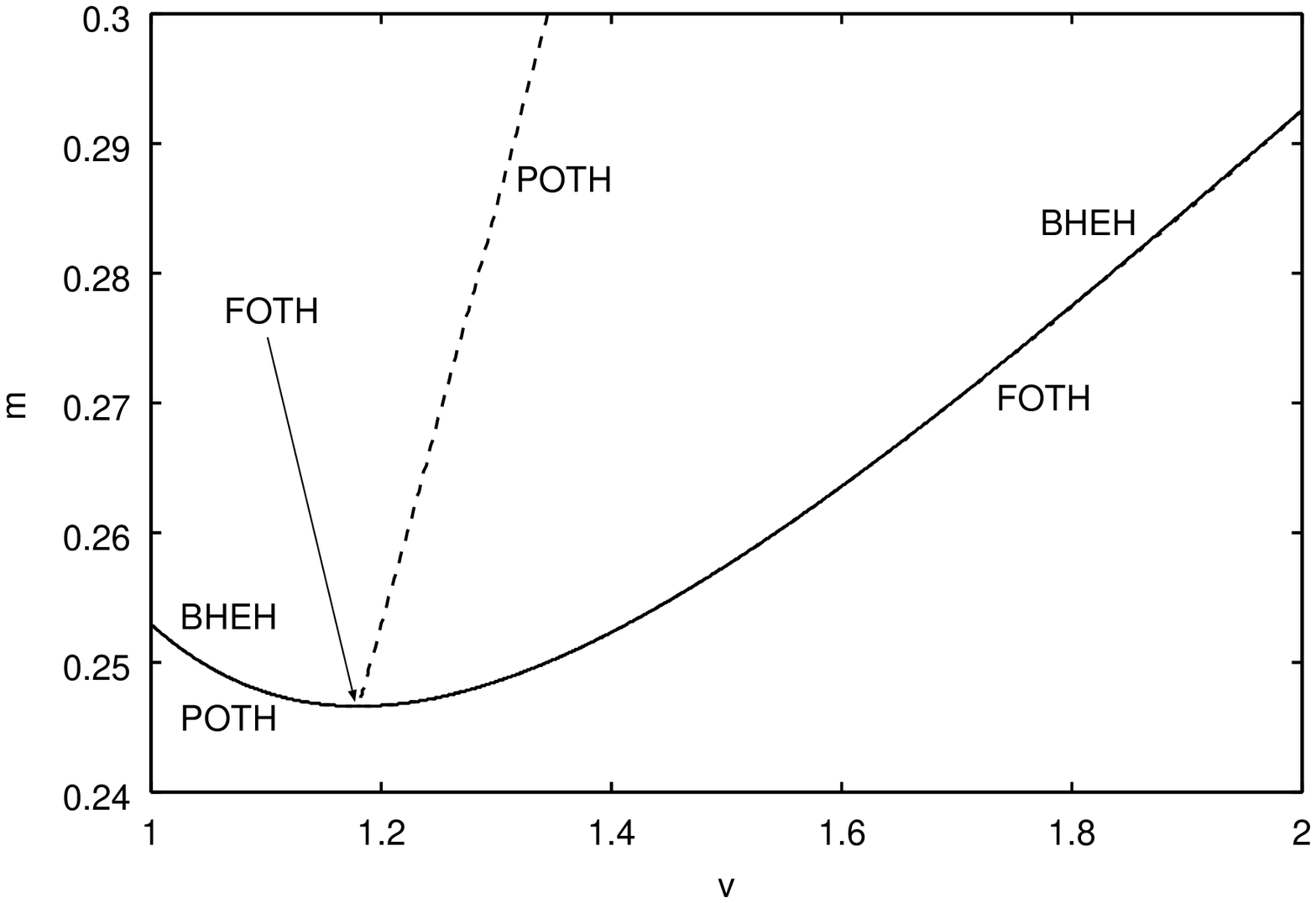}}
\end{tabular}
\caption{\label{fig:eh_mass}
The Misner-Sharp masses for the black hole event horizon (BHEH) and
trapping horizons with $r_{,v}=0$
and $r_{,u}=0$, plotted with solid, dotted and
dashed lines, respectively. 
The mass of the black hole event horizon  
decreases when it is in a past trapped 
region and increases when it is 
in a untrapped region with null expansions $(+,-)$.
The mass of the trapping horizon with $r_{,v}=0$ is 
slightly larger than that of the black hole event horizon
for each model, although they are almost indistinguishable.}
\end{figure*}
\end{center}

Figure~\ref{fig:eh_mass} shows the evolution of 
the Misner-Sharp mass of the BHEH and trapping horizons. 
For models where the initial BHEH is larger than 
the Friedmann cosmological apparent horizon,
the mass of the BHEH
first decreases and then increases.
As we will see later, the mass of the BHEH decreases 
if and only if it is in the past trapped region.
When the black hole gets out of the past trapped region,
its mass starts to increase.
The mass of the trapping horizon with $r_{,v}=0$ is 
only slightly smaller than that of the BHEH, 
despite the radii being considerably different.
This is because the density inside the 
perturbed region is very small.
After the BHEH crosses the trapping horizon with $r_{,u}=0$, which 
is a POTH at the crossing, 
the mass of the BHEH soon gets much smaller than the mass of the 
cosmological apparent horizon.

\begin{center}
\begin{figure*}[htbp]
\includegraphics[scale=1]{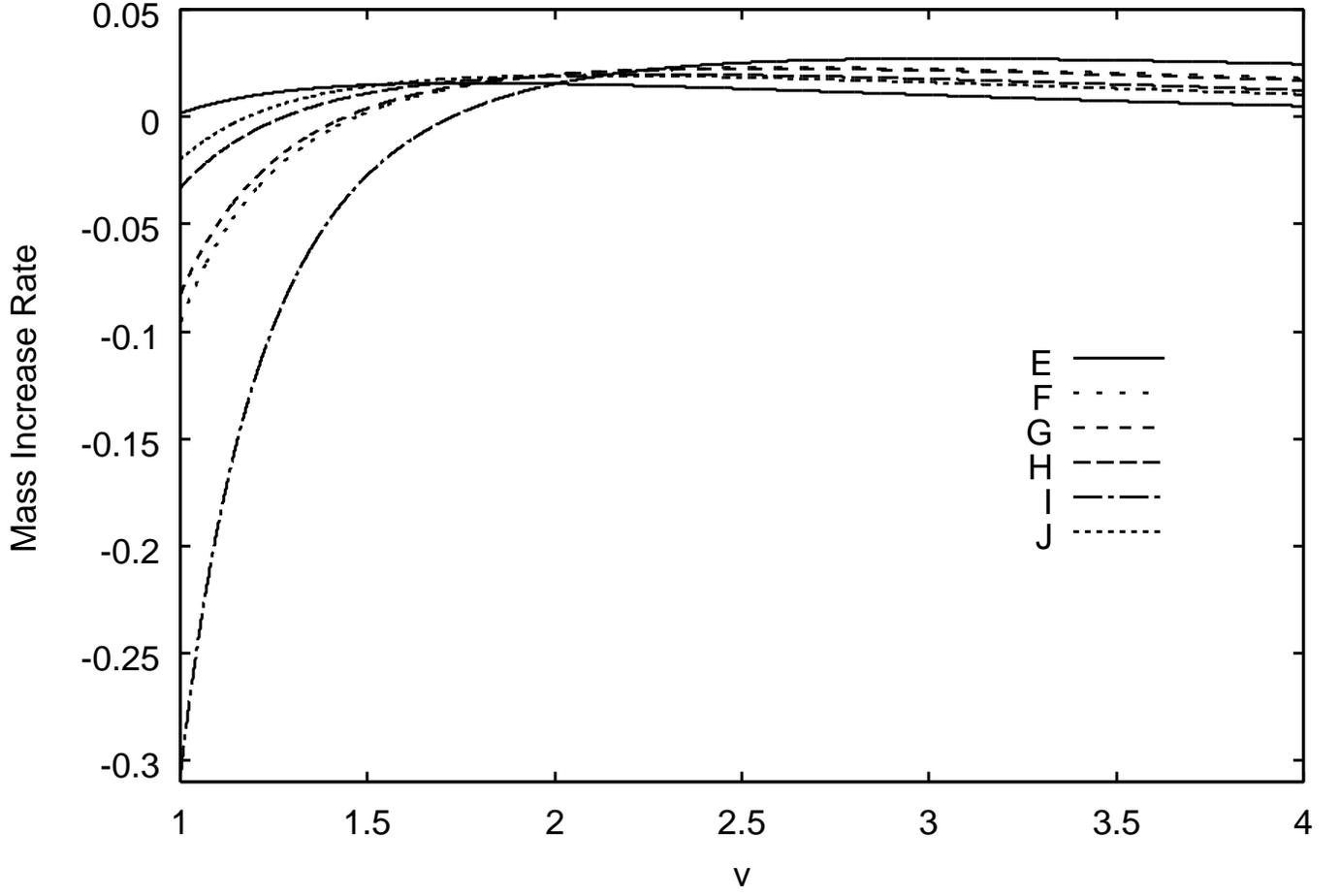}
\caption{\label{fig:mass_accretion}
Dependence of mass increase rate $dm_{\rm BHEH}/dv$
of the black hole event horizon on $v$
for Models E--J.
It is negative
when the black hole event horizon is in the past trapped region 
but positive when it gets out 
of the past trapped region.
}
\end{figure*}
\end{center}

The change in the BHEH mass may be 
seen more clearly in Fig.~\ref{fig:mass_accretion}, which shows
the rate of mass increase $dm_{\rm BHEH}/dv$. 
For models other than E, 
this is initially 
negative but it then increases, crosses zero, reaches a
positive maximum and then decreases.  
The larger the initial mass ratio of the BHEH
to the Friedmann cosmological apparent horizon, the 
more negative the initial
mass increase rate is.
For Model E, where the initial BHEH
is slightly smaller than the Friedmann cosmological 
apparent horizon, 
the accretion rate starts with a very small positive value,
increases to a maximum and then decreases.
The behaviour of the BHEH mass is clearly explained
by Eq.~(\ref{eq:m_v}), which can be rewritten 
as a black hole mass equation:
\begin{equation}
\frac{dm_{\rm BHEH}}{dt}=-8\pi r^{2} \frac{(\Psi_{,v})^{2}}{a^{2}}
\left(\frac{dr}{dt}\right)_{v=\mbox{const}}, 
\label{eq:accretion_t}
\end{equation}
where $t=t(u+v)$ is any time coordinate which depends on $u+v$.
Whether the black hole mass increases or decreases
depends solely on the sign of $r_{,u}$, which is the null
expansion along the $v=\mbox{const}$ direction. In the usual situation, 
the BHEH is in a region where $r_{,u}<0$
and its mass monotonically increases.
However, when the BHEH is in a past trapped region, 
its mass monotonically decreases because $r_{,u}>0$.
Equation~(\ref{eq:accretion_t}) (or (\ref{eq:m_v})) also explains why 
the mass accretion rate 
starts very small for Model E.
Since the BHEH is inside but 
very close to the cosmological apparent horizon, 
$r_{,u}$ starts off negative but very close to zero.
After some evolution, $r_{,u}$ decreases well below zero
and the accretion then increases.
This suppression of accretion for a PBH as large as a 
cosmological apparent horizon is due purely to general 
relativistic effects, or relativistic cosmological expansion.
Paper I discusses the qualitative difference between 
PBHs whose size is comparable to and much smaller 
than the cosmological apparent horizon.

For models with the initial BHEH larger than 
the Friedmann cosmological apparent horizon, the areas and masses of both 
trapping horizons decrease as $v$ increases 
before they cross each other.
After the crossing, they increase.
This behaviour is completely consistent with the ``second law''
for trapping horizons, as formulated by Hayward~\cite{hayward1996}. 
The theorem states, for example, that
if the null energy condition holds, then the area and mass of the 
FOTH (POTH) with $r_{,v}=0$ do not decrease (increase) along the vector $z$ 
tangent to the horizon. This vector has the form
$z=\beta \partial_{v}-\alpha \partial _{u}$ where $\beta>0$. 

\subsection{Conversion from a white hole to a black hole}
\begin{center}
\begin{figure*}[htbp]
\begin{tabular}{cc}
\subfigure[E]{\includegraphics[scale=0.5]{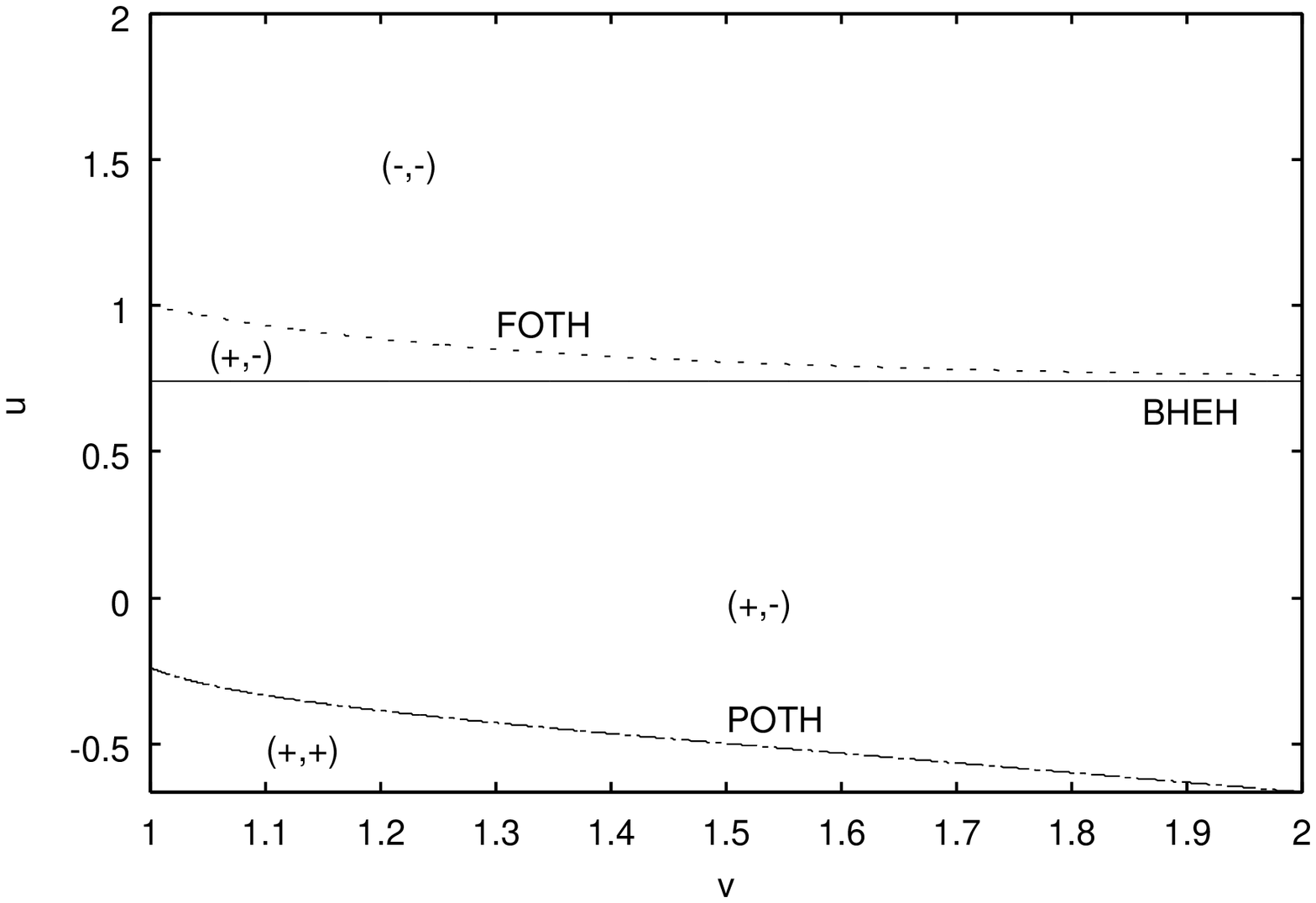}}&
\subfigure[F]{\includegraphics[scale=0.5]{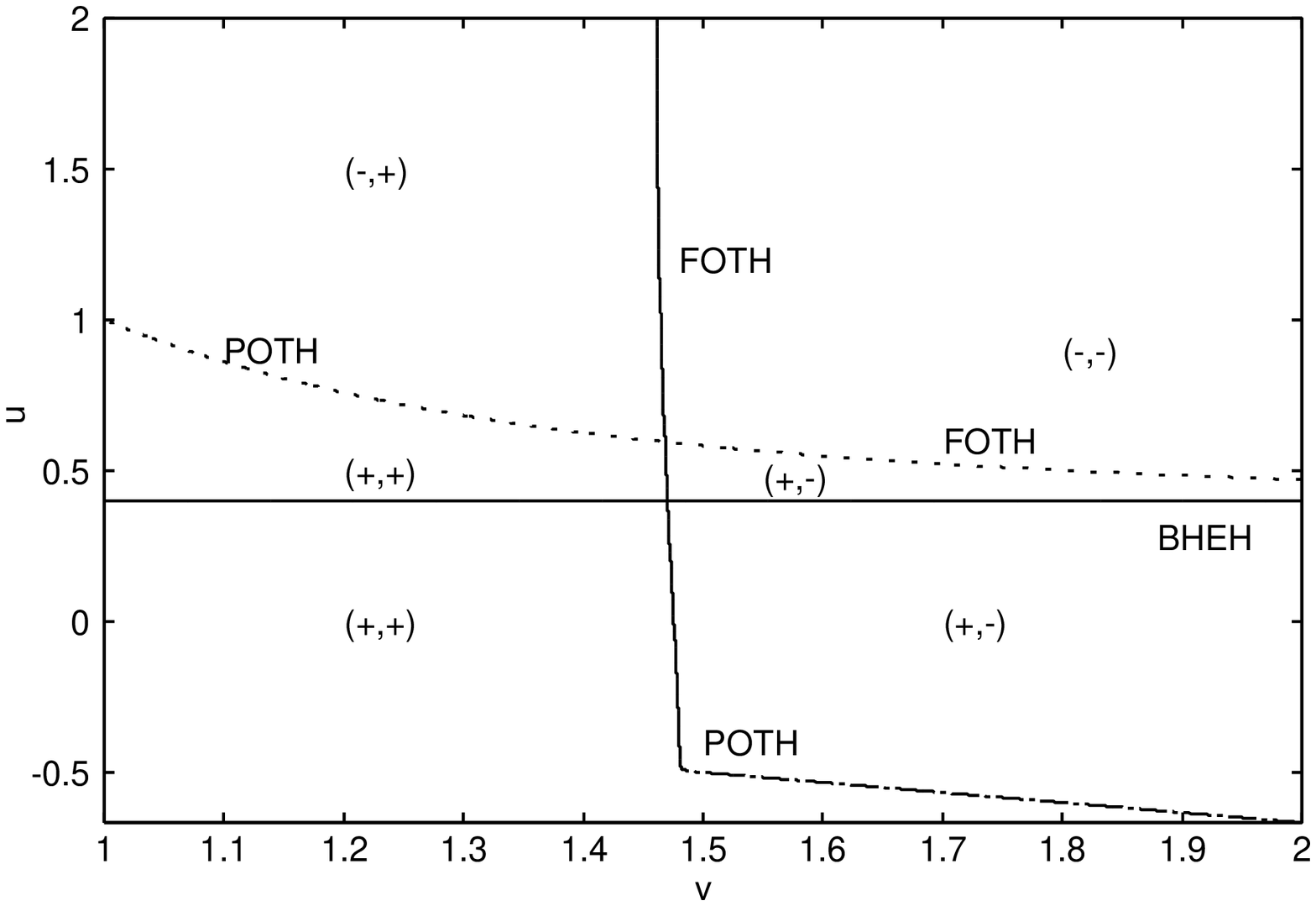}}\\
\subfigure[G]{\includegraphics[scale=0.5]{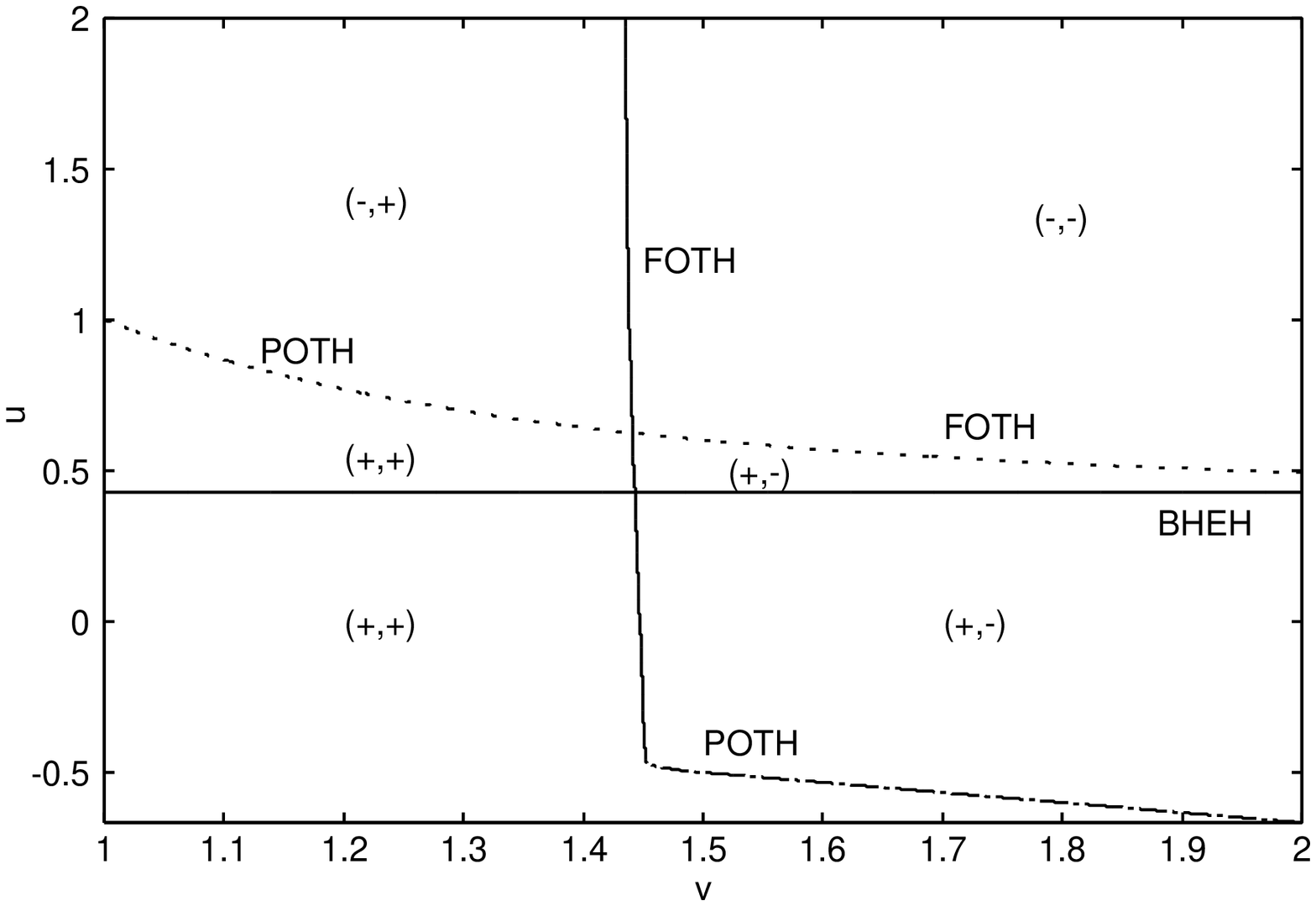}}&
\subfigure[H]{\includegraphics[scale=0.5]{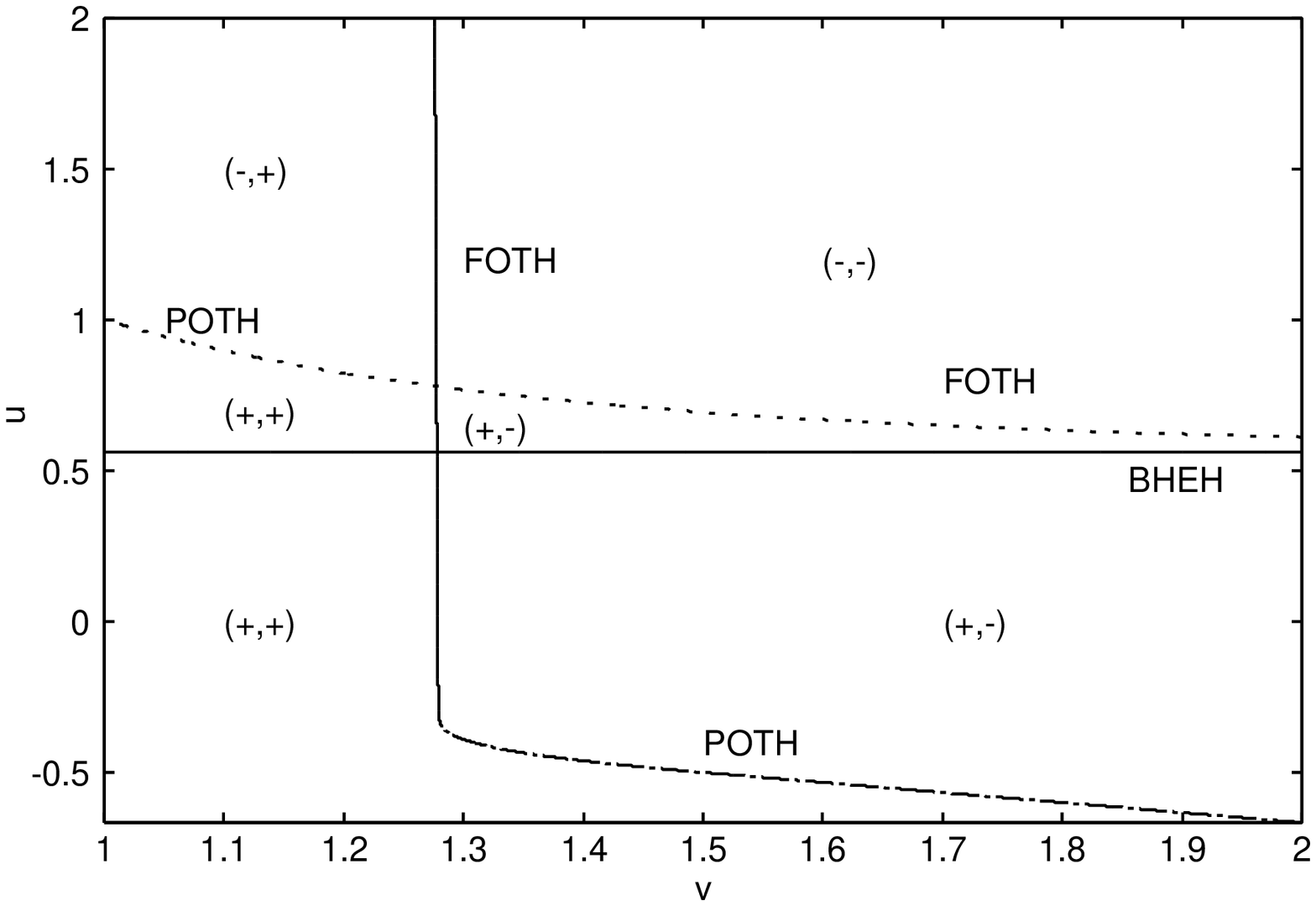}}\\
\subfigure[I]{\includegraphics[scale=0.5]{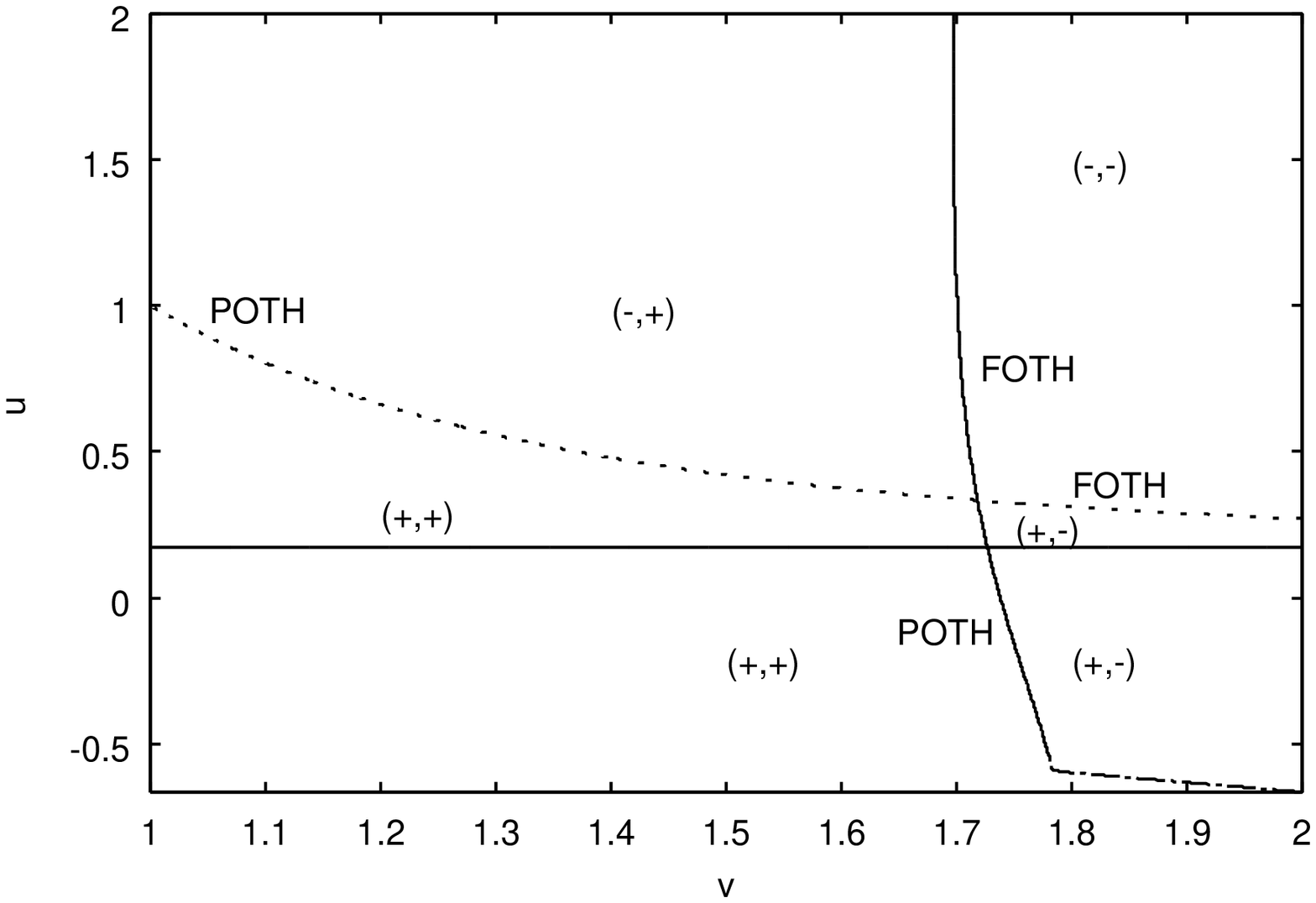}}&
\subfigure[J]{\includegraphics[scale=0.5]{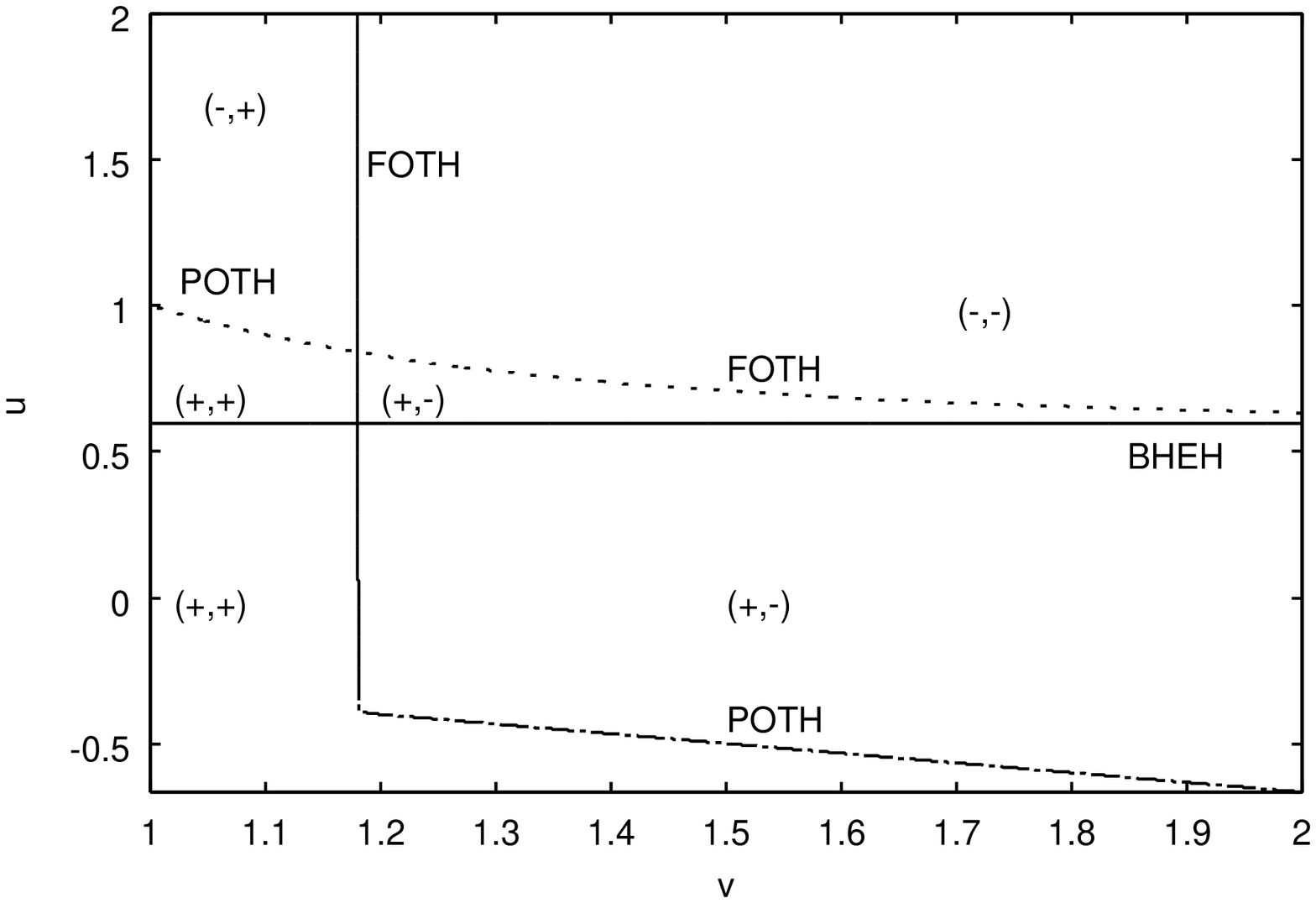}}
\end{tabular}
\caption{\label{fig:horizons_uv} 
Positions in $(u,v)$ plane of black hole event horizon (BHEH) and 
trapping horizons with $r_{,v}=0$ and $r_{,u}=0$, plotted
with solid, dashed and dotted-dashed lines, respectively,
The future and past outer trapping horizons are labelled as
``FOTH'' and ``POTH'', respectively.
The signs of $(r_{,v})$ and $(r_{,u})$ are also 
shown. A region is future trapped if it has
 $(-,-)$, past trapped if it has $(+,+)$
and untrapped if it has $(+,-)$ or $(-,+)$.}
\end{figure*}
\end{center}

\begin{figure*}[htbp]
\begin{center}
\begin{tabular}{cc}
\subfigure[]{\includegraphics[scale=0.5]{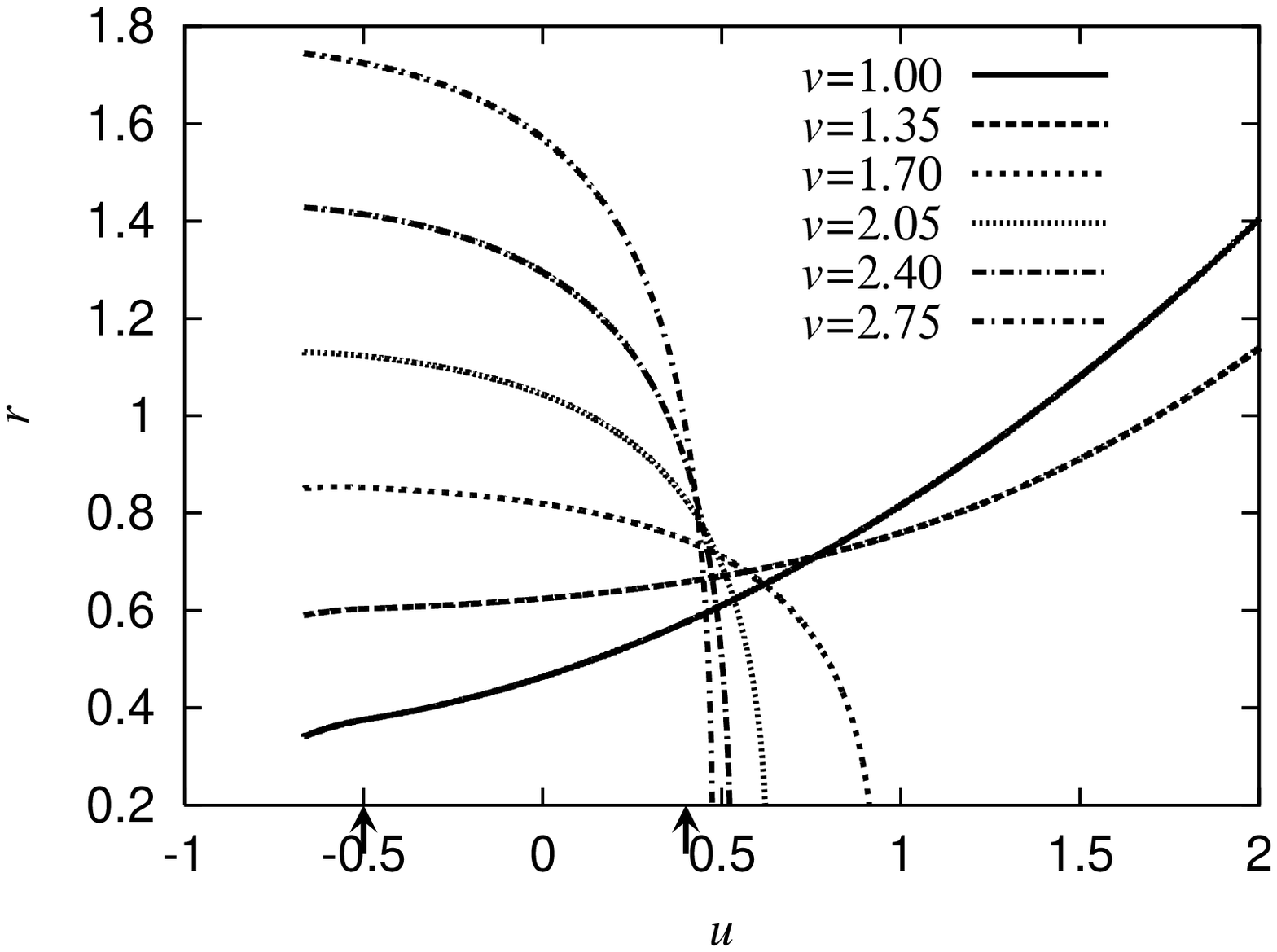}}&
\subfigure[]{\includegraphics[scale=0.5]{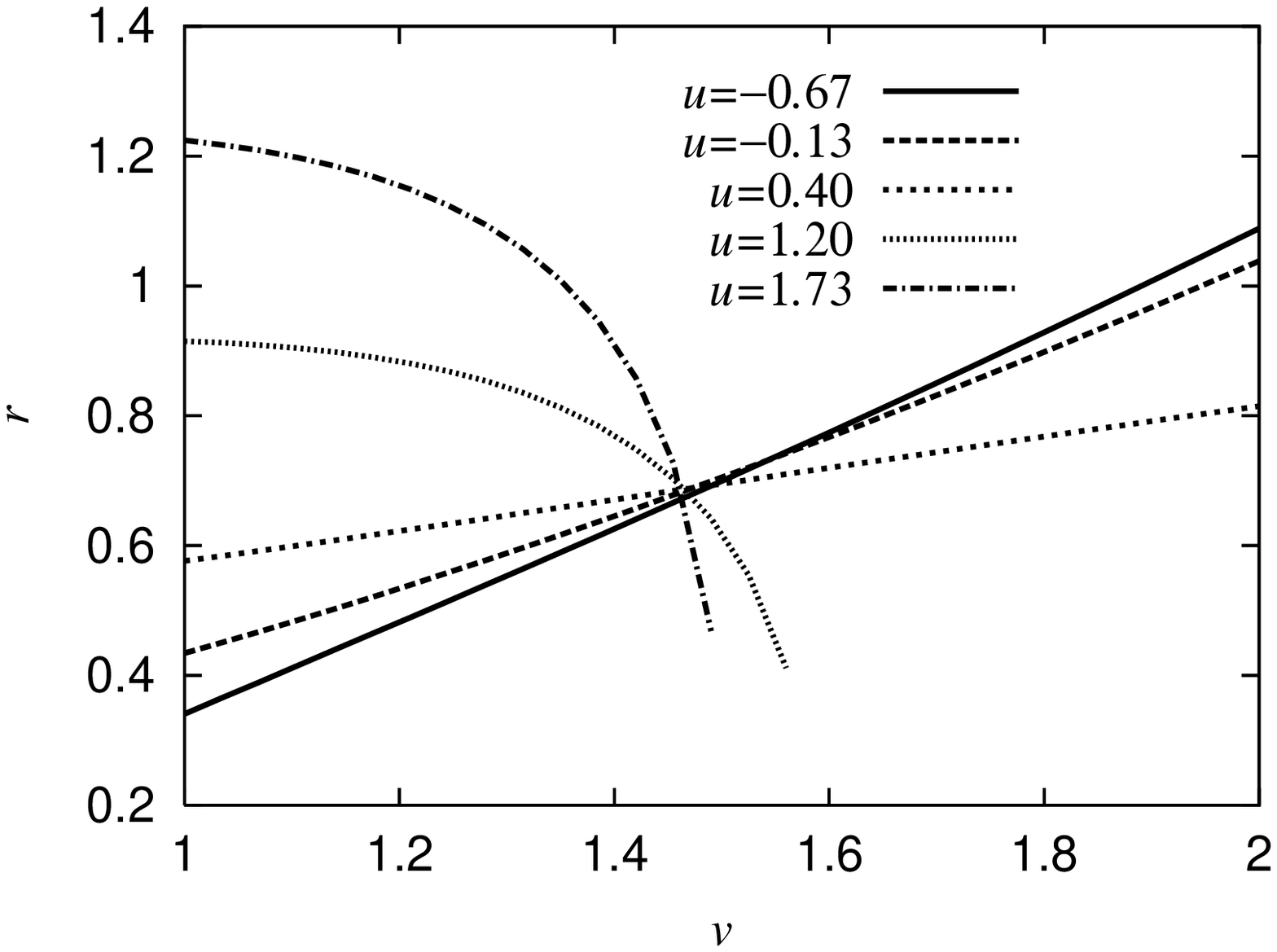}}\\
\subfigure[]{\includegraphics[scale=0.5]{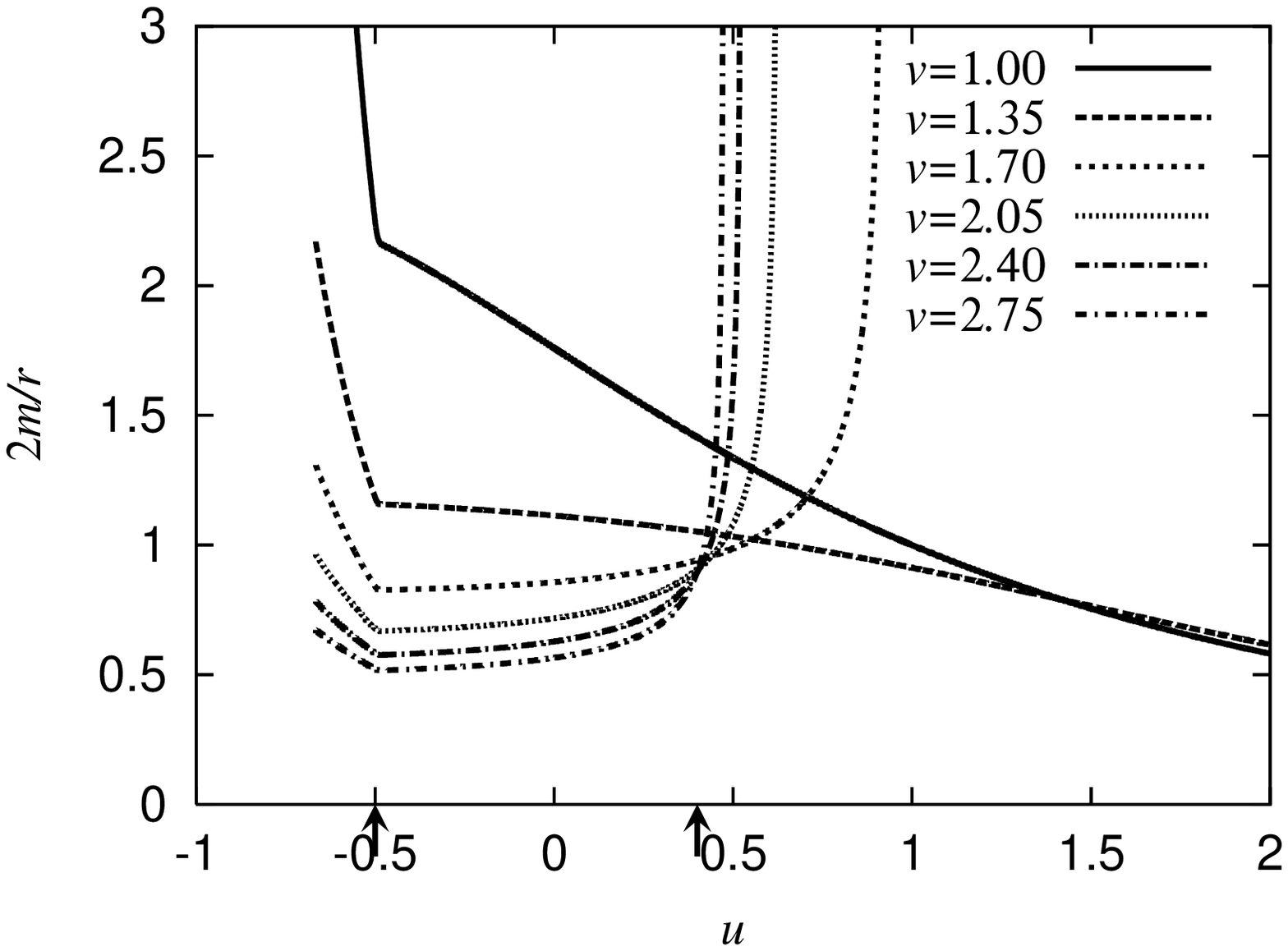}}&
\subfigure[]{\includegraphics[scale=0.5]{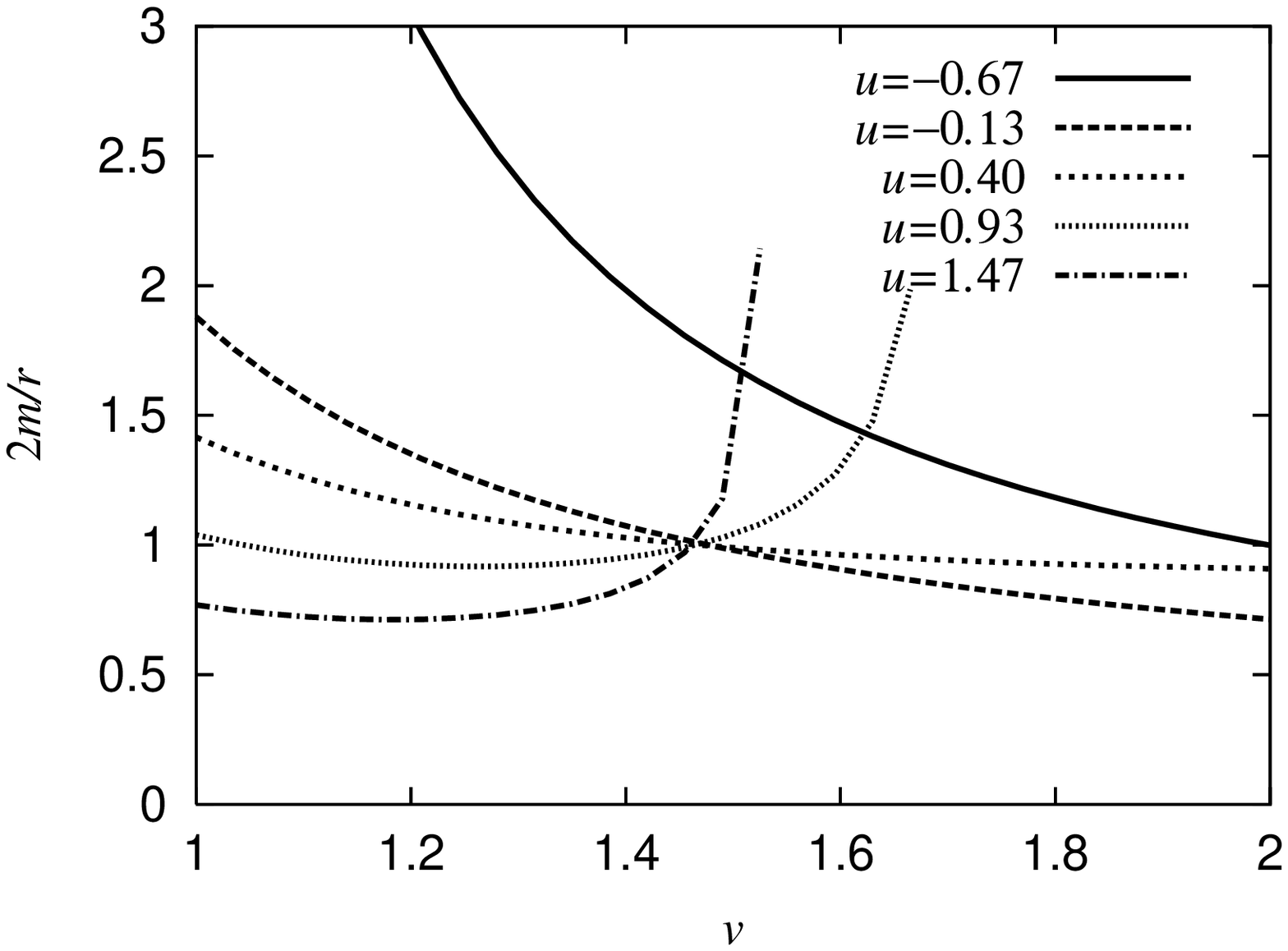}}
\end{tabular}
\end{center}
\caption{\label{fig:model_e_detail}
The evolution of $r$ along (a) $v=\mbox{const}$ 
and (b) $u=\mbox{const}$ and of $2m/r$ along (c) $v=\mbox{const}$ and (d) 
$u=\mbox{const}$ for Model F. 
In (a) and (c), the arrow at
 $u=-0.5$ denotes the matching surface, while the arrow at $u\simeq 0.400$
denotes the black hole event horizon.}
\end{figure*}

To understand the spacetime structure for solutions
in which the PBH is larger than the cosmological apparent horizon,
we concentrate on Model F. Figure~\ref{fig:model_e_detail}
gives the detailed numerical results for this model.
Figures~\ref{fig:model_e_detail}(a) and (b) give the evolution of $r$ 
along the null geodesics with 
$v=\mbox{const}$ and $u=\mbox{const}$,
respectively, while Figs.~\ref{fig:model_e_detail} (c) and (d)
give that of $2m/r$ along the null geodesics 
with $v=\mbox{const}$ and $u=\mbox{const}$, respectively.
It is seen that $r$ continues 
to increase along both the earlier null geodesics with 
$u=\mbox{const}$ and $v=\mbox{const}$.
This is also consistent with Fig.~\ref{fig:horizons_uv} (a).
On the initial null ray $v=v_{0}=1$,
$2m/r$ decreases monotonically as $u$
increases and crosses unity from above,
as can be seen in Fig.~\ref{fig:horizons_uv}(c).
Since we have chosen $\Psi=\mbox{const}$ for 
$u\geq u_{\rm m}+\Delta u$ on the initial null ray $v=v_{0}$,
Eq.~(\ref{eq:m_u}) implies $m=\mbox{const}$ here.
Since $r_{,u}=0$ implies $2m/r=1$, 
the area radius along the null ray $v=v_{0}$
must monotonically increase as $u$ increases.
This means that this null ray does not reach $r=0$
but another null infinity, different from the one which  
the null rays with $u=\mbox{const}$ reach.

\begin{figure}[htbp]
\includegraphics[scale=1]{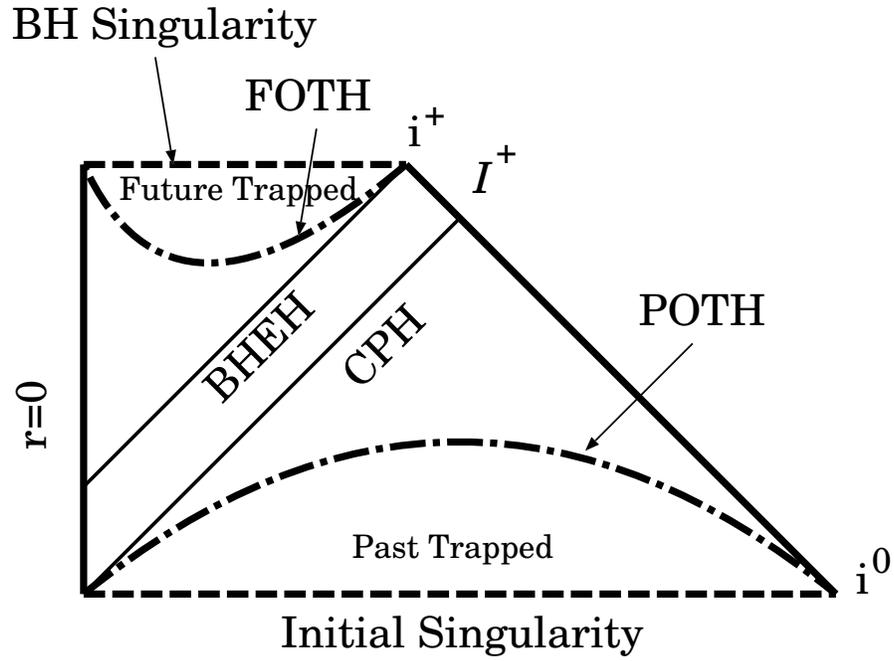}
\caption{\label{fig:pbh_clean}
The conformal diagram for a PBH smaller than 
the Friedmann cosmological particle horizon.}
\end{figure}

\begin{figure}[htbp]
\includegraphics[scale=0.8]{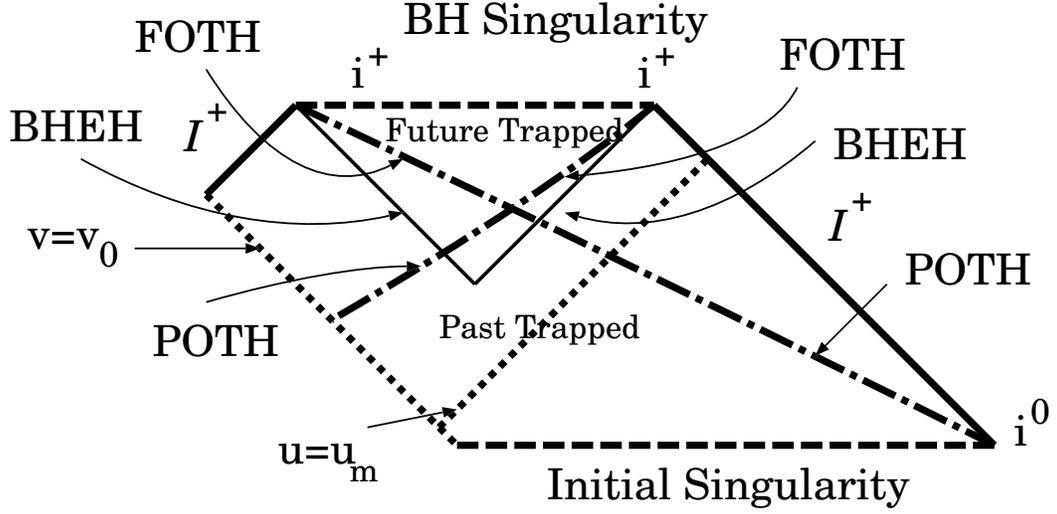}
\caption{\label{fig:pbh_big}
Conformal diagram showing the possible causal structure for
a PBH larger than the Friedmann cosmological apparent horizon.
There are two distinct future null infinities and a past trapped (white hole)
region is converted into a future trapped (black hole) region.
The region with  $u<u_{\rm m}$ is the usual flat 
 Friedmann spacetime but the region $v<v_{0}$ is not calculated. See text for details.}
\end{figure}

We can now see that the standard conformal diagram
for PBHs, which is given by
Fig.~\ref{fig:pbh_clean}, will not apply if
the BHEH is larger than the cosmological apparent horizon.
Rather 
the conformal diagram is given by 
Fig.~\ref{fig:pbh_big}.
The FOTH and POTH intersect
at one point in this diagram. This point corresponds to a sphere which is a 
bifurcating marginal surface.
It is clear that the spacetime can be interpreted as the 
conversion of a white hole to a black hole, since a PBH larger than the cosmological 
apparent horizon involves a past trapped region changing to a future 
trapped one.
There are two distinct asymptotic regions: the future null infinity for 
rays with $u=\mbox{const}$
and the one for rays with $v=\mbox{const}$. These two asymptotic
regions are associated with two disconnected untrapped regions. 
It is impossible for an observer to go from one 
untrapped region to the other by traversing the crossing point, the
future trapped or past trapped region.
The spacetime has no regular centre.

\begin{figure}[htbp]
\includegraphics[scale=1]{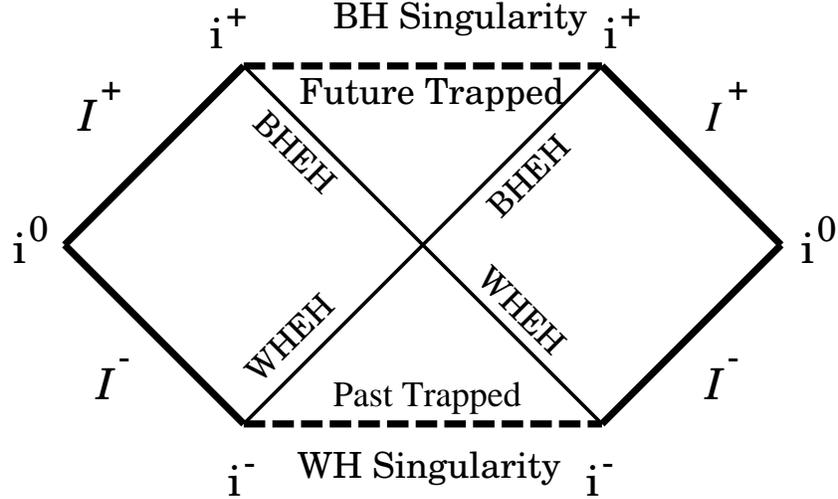}
\caption{\label{fig:schwarzschild_kruskal}
Conformal diagram of the maximally extended Schwarzschild
spacetime, i.e., the Kruskal diagram. 
The black hole event horizon (BHEH) 
coincides with the future outer trapping horizon
and the white hole event horizon (WHEH) coincides with the 
past outer trapping horizon.}
\end{figure}

Figure~\ref{fig:pbh_big} is naturally interpreted as an amalgamation of the Kruskal diagram for the extended
Schwarzschild solution, shown in Fig.~\ref{fig:schwarzschild_kruskal}, and the
standard conformal diagram for a black hole in a flat Friedmann
background, shown in Fig.~\ref{fig:pbh_clean}. It will be recalled that
the Kruskal diagram also contains a white hole changing into a black hole 
and two asymptotically flat regions. The analogy of this extension for the
conformal diagram shown in Fig.~\ref{fig:pbh_clean} is shown in
Fig.~\ref{fig:pbh_big_complete}. This contains the big bang singularity, 
the black hole singularity and two asymptotically flat Friedmann
regions. There are two BHEHs and two POTHs, these crossing 
in the black hole region. 
The regions below the two POTHs are past trapped but 
not precisely equivalent to a white hole since, 
unlike the situation in Fig.~\ref{fig:schwarzschild_kruskal}, 
there are no past null infinities and no past event horizons. 
Since the spacetime has no regular centre,
it cannot contain a particle horizon as a light cone
which emanates from the regular centre just after the big bang.
Figure~\ref{fig:pbh_big} is obviously contained within 
Fig.~\ref{fig:pbh_big_complete} but the present 
numerical calculations
do not determine what happens in the region $v<v_0$. 

\section{Discussion}

\begin{figure}[htbp]
\includegraphics[scale=1]{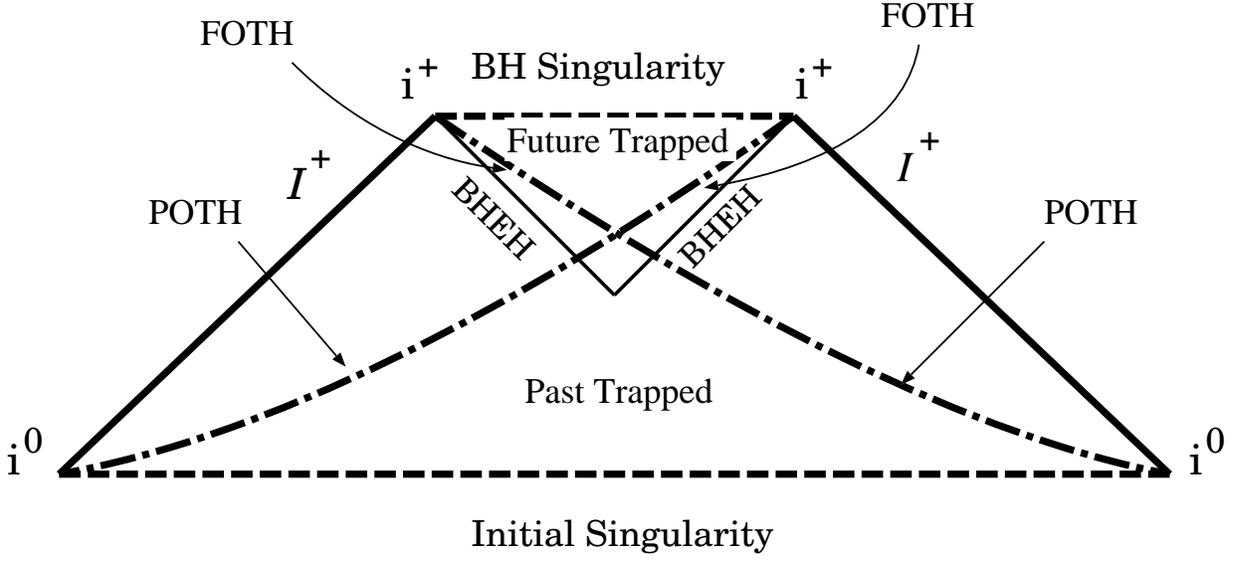}
\caption{\label{fig:pbh_big_complete}
Conformal diagram of a possible causal structure 
of the maximally extended spacetime for
a PBH larger than the Friedmann cosmological apparent horizon.
There are two distinct future null infinities, corresponding to the
conversion of a past trapped (white hole)
region into a future trapped (black hole) region.}
\end{figure}

If we accept both the conventional conformal diagram for PBHs 
shown in Fig.~\ref{fig:pbh_clean} and the new
one shown in Fig.~\ref{fig:pbh_big} or
Fig.~\ref{fig:pbh_big_complete},
we have two distinct PBH causal structures.
The conventional diagram has a regular centre $r=0$ and a single
null infinity, while the new one has no regular centre and 
two null infinities. This raises the issue of what determines 
the causal structure of PBH spacetimes. 
If we consider standard PBH formation
from initial data with a regular centre, then the causal structure
will be described by the conventional diagram.

\begin{figure}[htbp]
\includegraphics[scale=1]{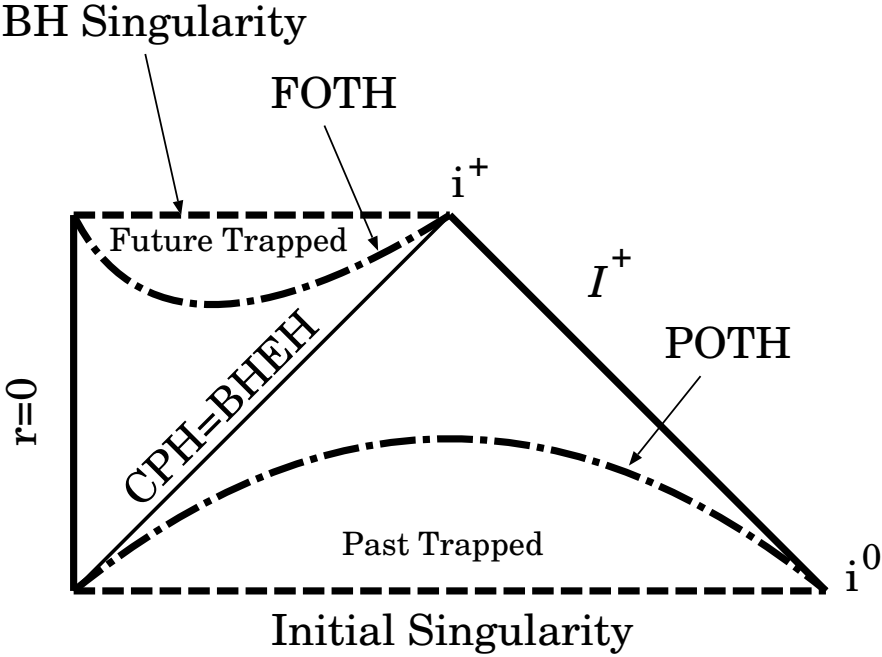}
\caption{\label{fig:pbh_transition}
Conformal diagram 
of the critical PBH spacetime, where the black hole
event horizon (BHEH) coincides with the 
cosmological particle horizon (CPH).}

\end{figure}

Another issue concerns the transition 
between the two diagrams. If this transition is governed 
by parameters which can be 
changed continuously, it would be natural to assume that there is a
threshold spacetime which separates them.
We should note that the BHEH is null, while the cosmological 
apparent horizon is spacelike in the massless scalar field case
(see Fig.~\ref{fig:local_perturbation}).
Hence, if the $u_{\rm BHEH}$ is smaller than 
$u=u_{\rm CPH}$ of the cosmological particle horizon 
($0$ in the present coordinates),
the BHEH can be always outside the cosmological apparent horizon
if we take $v_{0}$ sufficiently small.
In this case, $m_{\rm BHEH}(v_{0})>m_{\rm CAH}(v_{0})$ 
and hence the causal structure should be given by
Fig.~\ref{fig:pbh_big} or Fig.~\ref{fig:pbh_big_complete}.
On the other hand, if $u_{\rm BHEH}>u_{\rm CPH}$, the causal structure
is depicted by Fig.~\ref{fig:pbh_clean} and the spacetime 
has a regular centre before the BHEH appears.
Therefore, the transition spacetime corresponds to a PBH whose 
event horizon coincides with the cosmological particle
horizon of the background cosmological solution.
The causal structure of this critical
spacetime is depicted in Fig.~\ref{fig:pbh_transition}.

The discussion of the mass variation of the BHEH 
can be understood in the more general spherically symmetric context
from the counterparts of Eqs.~(\ref{eq:m_u}) and (\ref{eq:m_v})~\cite{hayward1996,hc2004a}. 
In our notation, these equations become
\begin{eqnarray}
m_{,u}&=&\frac{8 \pi r^{2}}{a^{2}}(T_{uv}r_{,u}-T_{uu}r_{,v}), \\
\label{eq:m_u_general}
m_{,v}&=&\frac{8 \pi r^{2}}{a^{2}}(T_{uv}r_{,v}-T_{vv}r_{,u}).
\label{eq:m_v_general}
\end{eqnarray}
The combination of the two terms in parentheses on the right-hand side 
of Eq.~(\ref{eq:m_v_general}), related to null expansions,
determines the time variation of the BHEH mass.
In the case of a massless scalar field, the situation is simplified 
because
$T_{uv}=0$
and $T_{vv}=(\Psi_{,v})^{2}\geq 0$.
We therefore conclude that 
the black hole accretion vanishes when the event horizon coincides with
the POTH with $r_{,u}=0$ 
and that it becomes negative when the event horizon is in a 
past trapped region.
For general matter fields, 
it can be proved that the BHEH mass 
is non-decreasing if $r_{,v}>0$, $r_{,u}<0$ and 
the dominant energy condition hold
on the event horizon (c.f. Proposition 5 of \cite{hayward1996}).
However, if these assumptions are not satisfied,
the mass variation of the BHEH is non-trivial.
If the dominant energy condition holds, then 
the conditions $T_{uv}\ge 0$ and $T_{vv}\ge 0$ are necessarily satisfied. 
In the general case with $T_{uv}>0$, 
the BHEH mass still increases even 
when it is on a POTH with $r_{,u}=0$.
When the BHEH is in a past trapped region, 
the first and second terms 
in parentheses on the right-hand side of Eq.~(\ref{eq:m_v_general})
make positive and negative contributions, respectively, 
to the black hole mass change.
Even when the BHEH is in a past trapped region, its mass can 
still increase if the first term is dominant.
This suggests that the BHEH has to be well outside the 
POTH in order to have suppressed accretion
or mass decrease for general matter fields satisfying
the dominant energy conditions.
On the other hand, there could still be 
a lot of accretion for some other kinds of matter field.

We have shown that the mass of the BHEH can decrease even if 
all possible energy and causality conditions hold.  
This suggests that the usual concept of event horizons 
is not appropriate for a proper discussion of black hole thermodynamics.
This is why Hayward~\cite{hayward1996,hayward1993,hayward1994,hayward1998} 
introduced the concept of the FOTH (POTH)
as a useful generalisation of the idea of a black hole 
(white hole) event horizon.  
As we saw in Section~\ref{sec:mass_horizon},
the idea is that all definitions should be quasi-local and this is useful in proving black hole properties.
For example, when the black hole (white hole) horizon is
defined as a FOTH (POTH),
one can prove the monotonicity of the area and mass in the 
spherically symmetric situation.
From this point of view, Fig.~\ref{fig:pbh_big} 
describes the conversion of a white hole to a black hole by definition. 

Although the analysis in this paper has assumed spherical
symmetry, it is likely that similar considerations apply more generally.
If we consider slightly nonspherical PBH spacetimes, 
it is trivial to show that the strange properties discussed in this 
paper still hold. However, when the system is very far from spherical symmetry,
we need to re-analyse the Einstein equations.
Although the conditions for suppressed accretion or 
mass reduction may change,
we believe the appearance of these properties 
is robust.

Since a spacetime which contains a PBH 
larger than the cosmological apparent horizon cannot
have a regular centre, such PBHs could not form 
through classical processes from a Friedmann background.
This implies that a PBH formed through classical processes 
always increases its mass and soon gets much smaller than
the cosmological horizon even if it is comparable with the
cosmological scale at its formation.
On the other hand, if we take quantum processes or 
inflation preceding to the 
massless scalar field dominated stage into account,
the formation of such PBHs may be allowed. 
Moreover, there is a possibility that the universe 
initially contained a past trapped region which was
converted into a future trapped region, 
as indicated in Fig.~\ref{fig:pbh_big_complete}.
This kind of PBH might be described as ``ab initio'' or ``eternal''.

\section{Conclusion}
We have investigated PBHs in a flat Friedmann universe
with a massless scalar field which are larger than the cosmological 
apparent horizon.
Through fully general relativistic numerical calculations,
we have found that there are
two trapping horizons which cross. This corresponds to
the conversion of a white hole into a black hole
and to an initial reduction in the mass of the black hole event horizon.
We have seen that these unusual features are peculiar 
to PBHs larger than the cosmological apparent horizon.

The present analysis is confined to the 
spherically symmetric situation with a massless scalar field,
so it is clearly important to examine how sensitive
these peculiar features are 
to the assumed symmetry and type of matter field.
In particular, it is interesting to study whether
the condition for the black hole mass reduction
applies more generally. 
If so, this will be important for
black hole physics and
black hole thermodynamics.
It is also interesting to study under what circumstances
such a PBH can arise in the early universe, 
especially in the context of inflationary cosmology, 
and how this relates to the separate universe condition. 
To make such an extension numerically or analytically, 
it would be very important
to adopt the appropriate definition of the cosmological and black hole 
horizons and the trapping horizon framework should be suitable
for the case in which the black hole is as large as 
the cosmological horizon.
We will study these issues in a future paper.

The formation of PBHs arising from the evolution of 
an effectively massless scalar field during inflation and 
its implication for the formation of 
galaxies and supermassive black holes in galactic nuclei
has been recently discussed in Ref.~\cite{rsk2001}.
In this scenario, super-horizon PBHs may naturally 
arise due to the closed domain walls of size exceeding the
cosmological horizon.

Finally, it should be noted that the conversion of a wormhole to a black hole
in a first order phase transition was studied in Ref.~\cite{kssm1981} in the
context of bubble nucleation. They showed that if a wormhole is created
in the transition from a false to a true vacuum, it necessarily 
collapses to a black hole. Since our system only contains a
scalar field with no potential and therefore no phase transition,
the situation is rather different. However, 
it is quite interesting that similar
phenomena are seen in different systems.

\acknowledgments
We would thank S.~A.~Hayward for 
helpful comments and clarifying trapping horizon notation.
TH is grateful to K.~Nakao for helpful comments.
TH was partly supported from JSPS.
This work was partly supported by the Grant-in-Aid for the 21st Century 
COE ``Center for Diversity and Universality in Physics'' 
from the Ministry of Education, Culture, Sports, 
Science and Technology (MEXT) of Japan.

\newpage

\newpage

\end{document}